\documentclass[12pt,a4paper,oneside]{article}
\usepackage{graphicx,amssymb,amsmath,slashed,cite}
\usepackage[margin=2cm]{geometry}

\newcommand{\beq}{\begin{equation}}
\newcommand{\eeq}{\end{equation}}
\newcommand{\bea}{\begin{eqnarray}}
\newcommand{\eea}{\end{eqnarray}}
\newcommand{\co}{\mathcal{O}}

\newcommand{\tr}{\mathrm{Tr}}

\begin{document}
\topmargin -1.8cm
\oddsidemargin -0.8cm
\evensidemargin -0.8cm

\thispagestyle{empty}

{\tiny CERN-PH-TH/2012-201}

\vspace{20pt}

\begin{center}
\vspace{20pt}

\Large \textbf{Up Asymmetries From\\ Exhilarated Composite Flavor Structures}

\end{center}

\vspace{15pt}
\begin{center}
{\large  Leandro Da Rold$^{\, a}$, C\'edric Delaunay$^{\, b}$, Christophe Grojean$^{\, b}$, Gilad Perez$^{\, b,c}$}

\vspace{20pt}

$^a$\textit{Centro At\'{o}mico Bariloche and Instituto Balseiro, 8400 San Carlos de Bariloche, Argentina}\\[0.2cm]
$^b$\textit{CERN Physics Department, Theory Division, CH-1211 Geneva 23, Switzerland}\\[0.2cm]
$^c$\textit{Department of Particle Physics and Astrophysics, Weizmann Institute of Science,\\ Rehovot 76100, Israel}

\end{center}

\vspace{20pt}
\begin{center}
\textbf{Abstract}
\end{center}
\vspace{5pt} {\small \noindent

We present a class of warped extra dimension (composite Higgs) models which conjointly accommodates the $t\bar t$ forward-backward asymmetry observed at the Tevatron and the direct CP asymmetry in singly Cabibbo suppressed $D$ decays first reported by the LHCb collaboration. We argue that both asymmetries, if arising dominantly from new physics beyond the Standard Model, hint for a flavor paradigm within partial compositeness models in which the right-handed quarks of the first two generations are not elementary fields but rather composite objects. We show that this class of models is consistent with current data on flavor and CP violating physics, electroweak precision observables, dijet and top pair resonance searches at hadron colliders. These models have several predictions which will be tested in forthcoming experiments.
The CP asymmetry in $D$ decays is induced through an effective operator of the form $(\bar u c)_{V+A}(\bar s s)_{V+A}$ at the charm scale, which implies a larger CP asymmetry in the $D^0\to K^+K^-$ rate relative the $D^0\to \pi^+\pi^-$ channel. This prediction is distinctive from other Standard Model or dipole-based new physics interpretation of the LHCb result. CP violation in $D-\bar D$ mixing as well as an an excess of dijet production of the LHC are also predicted to be observed in a near future.
A large top asymmetry originates from the exchange of an axial resonance which dominantly produces left-handed top pairs. As a result a negative contribution to the lepton-based forward-backward asymmetry in $t\bar t$ production, as well as $\mathcal{O}(10\%)$ forward-backward asymmetry in $b\bar b$ production above $m_{b\bar b}\simeq 600\,$GeV at the Tevatron is expected. 
}

\vfill\eject
\noindent

\section{Introduction}
The two most fundamental
questions in particle physics are the origin of electroweak (EW) symmetry breaking and the stability of the EW scale far below the energy scale associated to gravitational interactions. It is now an exciting time where the LHC experiments are closing in on answering the former question in the form of a 125$\,$GeV Higgs boson~\cite{Htalk}.
The latter question is yet to be answered despite intense experimental efforts. An interesting theoretical speculation where the EW scale is naturally low is in models where the Higgs is not an elementary field but a composite object~\cite{PC,CHM}. 
The Higgs mass can moreover be naturally light if the Higgs field arises as a pseudo Goldstone boson (PGB)~\cite{HPGB}.
An equivalent description of this idea is realized in Randall--Sundrum (RS) models where the Higgs field is confined to the IR region of a warped extra dimension~\cite{RS,GW}.\footnote{We freely use in this letter the terminology of both the composite Higgs and warped extra dimension frameworks.} 

This interesting framework has some predictive power. It induces modifications in the Higgs couplings to fermions and EW gauge bosons relative the Standard Model (SM)~\cite{SILH} and it implies the  presence of new physics (NP) resonances~\cite{RES}, as well as top fermionic partners~\cite{topfriends}, still awaiting to be discovered. 
Yet very little is learned {\it a priori} about the flavor sector of (and beyond) the SM within the warped/composite framework.  A similar situation is encountered in supersymmetry where models that attempt to solve the hierarchy problem also have a limited  power in predicting the structure of the SM flavor parameters as well as that of the soft supersymmetry breaking terms. 

Existing studies on the flavor structure of RS models fall into three broad classes.
(I) {\it Anarchy}: The microscopic flavor parameters, namely five dimension (5D) fermion masses and Yukawa couplings, are structureless (anarchic).
 This is the most explored case since, besides being most general, it has a very attractive feature when considering flavor physics. This feature consists of an integral mechanism to generate flavor hierarchies~\cite{NeuGross,Huber,GhergPom}, while maintaining the anarchic nature of the fundamental flavor parameters. In this case the SM mass hierarchies are dictated by the relative degree of compositeness of the SM fermions. Heavy SM fermions like the top quark are thus interpreted as mostly composite objects, while lighter SM fermions are mostly elementary fields. The same integral mechanism also protects against large contributions to flavor changing neutral current (FCNC) processes through a GIM-like mechanism~\cite{HuberFV,APS}. However, this so-called RS--GIM mechanism is not perfect and overly large (CP violating) contributions in the down sector as well as electric dipole moments are induced~\cite{APS,CsakiFalkowskiWeiler}.
It is worth pointing out though that the SM flavor hierarchies in masses and mixings along with a sufficient RS--GIM suppression can be obtained in anarchic models where the hierarchy problem is only solved up to scale much lower than the Planck scale~\cite{LRS}.
Another mechanism based on extending the color gauge symmetry in the bulk to SU(3)$_L\times$SU(3)$_R$ was also proposed in Ref.~\cite{Bauer:2011ah} to suppress flavor changing contributions in the down sector.\footnote{We present a fully detailed  analysis of this proposal in Appendix~\ref{App_kaon}.}   
(II) {\it Alignment}: The microscopic flavor parameters are anarchic, so the SM flavor puzzle is solved in a generic way, but the 5D additional sources of flavor breaking are aligned with the 5D Yukawa interactions in a way that down-type flavor violation is suppressed~\cite{horizontalsym,shining}.
Note that if the lepton sector follows the same structure, neutrino flavor anarchy both in masses and mixings is expected~\cite{PR}.
(III) {\it Minimal flavor violation} (MFV): The microscopic flavor parameters are hierarchical and realize the 4D MFV selection rules~\cite{MFV,GMFV}. The SM flavor puzzle remains unsolved but the theory entertains a strong mechanism to suppress new sources of flavor breaking~\cite{RZ,5DMFV,FTRS,RedWei}.

There is another interesting fact about the warped/composite framework. NP contributions to the EW parameters are suppressed when the SM fermions are embedded in appropriate representations under the custodial symmetry~\cite{custo,AgasheDarold}. In particular,  if the SM fermion weak singlets are also singlet of the custodial group it was observed that EW precision tests (EWPTs) are very permissive, allowing for ultra composite first two generation quarks~\cite{FTRS,RedWei}.
This suggests a different theoretical approach where the fundamental flavor parameters are not anarchic but rather display some peculiar structure. 
Models of this kind were proposed in the context of class (III) and they share several possible implications. 
The most dramatic ones are related to the fact that 
 new contributions to flavor diagonal processes involve all three generations (as opposed to the generic prediction of the anarchic framework, where only the third generation is at play~\cite{APS,Agashe:2006hk}, since the first two generation quarks are mostly elementary fields). A drastic change in the phenomenology resides in an increase in the resonance production rate~\cite{FTRS,extraordinary,RedWei} and a potential sizable contribution to the $t\bar t$ forward backward asymmetry ($A_{\rm FB}$). However, note that flavor physics tends to be generically not very predictive in models of class (III) since the deviation from the SM prediction are suppressed in a manner similar to how the SM itself restrains its interaction from generating large contribution to FCNC processes.
 
In this work we present a new class of models, class (IV), with {\it exhilarated} flavor structures. Those models combine ingredients of classes (II) and (III) in the sense that contributions to various flavor changing processes are suppressed by some (implicit) mechanism of alignment and the anarchy assumption is broken, thus allowing the first two generation singlet fields to be composite. This new class of models has exciting implications for the collider phenomenology related to flavor diagonal processes. Moreover large contributions to FCNC processes are possible, specifically in the weak singlet sector due to the light generation quarks' compositeness.
Although no direct evidence of NP beyond the SM has been found so far, a few intriguing data have recently emerged which could be interpreted as indirect manifestations of NP. Of particular interest are (a) the anomalous  forward-backward asymmetry ($A_{\rm FB}$) in top pair production at Tevatron~\cite{AFBCDF,AFBCDFdiL,AFBD0}, and (b) the now solid evidence for CP violation (CPV) in singly Cabibbo suppressed decays of $D^0$ mesons (made of $c\bar u$ valence quarks)~\cite{lhcb,BelleICHEP,CDF10784,Aaltonen:2011se,Staric:2008rx,Aubert:2007if,HFAG}. 
It is interesting that the current data can be interpreted as weakly hinting for flavor models of class (IV).
First of all, classes (III) and (IV) are consistent with the Tevatron's results (a) since a large $A_{\rm FB}$ generically requires sizable NP couplings to both the top and the valence quarks~\cite{WISEFT}.
Second, class (II) and (IV) support the result (b) as the large measured value of the charm CP violating observable $\Delta a_{CP}$ would require large contributions to $c\to u$ transition amplitudes if it were to be interpreted as coming from new physics.\footnote{See Refs.~\cite{Golden:1989qx,DaCPSM} for temptative SM-like explanations.}
Finally, we find worth pointing out that classes (II), (III) and (IV) are all consistent with the recent observations of a large $\theta_{13}$ neutrino mixing angle~\cite{theta13}.

The main focus of this work is to interpret the data described in items (a) and (b) within the warped/composite Higgs framework. We choose to remain agnostic about the origin of the flavor parameters considered in this paper. Instead, we  present an ansatz belonging to the new class of models outlined above, which we shape to address the above data without conflicting with other flavor and CP violating processes. Yet we argue that, although we do not attempt to obtain a microscopic realization of this ansatz, the data outlined above does suggest an interesting underlying flavor structure. 

We show that the observations (a) and (b) are accommodated in models where the right-handed quarks are composite objects. This setup is unique for it provides a unified explanation for the observed charm CP and top forward-backward asymmetries. Moreover it differs radically from other NP interpretations of CPV in charm decays since the latter is explained through a four-fermion operator involving a charm, an up and  a pair of strange/anti-strange quark singlets (rather than through a chromomagnetic dipole operator as proposed  in supersymmetric models~\cite{NirKagan,SUSYdipoles} and warped/composite models~\cite{Rattazzi,DelKam}). This unique explanation is very predictive and has several experimental implications. In particular, we show that an excess of dijet events over the SM prediction, as well as CP violating $D-\bar D$ mixings and a large CP asymmetry in the $D\to K^+K^-$  rate are expected.  

The paper is organized as follows. In Section~\ref{section_model}, we present our flavor setup within the warped/composite framework. In Sections~\ref{AFB} and~\ref{section_CPV-LHCb} we demonstrate that this setup accomodates the large reported asymmetries in top pair production and charm decay, respectively. Constraints from other flavor violating processes  and dijet searches at the LHC are addressed in Sections~\ref{flavor} and~\ref{sec:dijets}, respectively. We discuss in Section~\ref{othersignal} other possible collider signatures of the advocated flavor scenario. We finally conclude in Section~\ref{conc}.

\section{The model}\label{section_model}
We describe here the particular setup of warped/composite models which addresses the top and charm data. We choose to present our analysis in the warped extra dimension framework, but similar results are obtainable in the 4D composite Higgs dual. 
We work in a slice of AdS$_5$ space-time, whose metric in conformal coordinates is
$ds^2=(kz)^{-2}\left(\eta_{\mu\nu}dx^\mu dx^\nu-dz^2\right)$
with $\eta_{\mu\nu}={\rm diag}(+---)$ and a curvature scale
$k\simeq 10^{19}\,$GeV, thus solving the hierarchy
problem up to the Planck scale. The slice is
bounded by two branes at $z=R\sim k^{-1}$ and
$z=R\,'\sim\,$TeV$^{-1}$ usually referred to as the UV and IR
branes, respectively~\cite{RS}.
In the EW sector we consider the SU(2)$_L\times$SU(2)$_R\times$U(1)$_X$ (custodial) gauge symmetry in the bulk. This symmetry is broken down to the SM gauge symmetry in the UV by boundary conditions, and to the diagonal subgroup by a $({\bf 2,2})_0$ bulk Higgs field, $H$, localized towards the IR~\cite{custo}. We consider a  Higgs vacuum expectation value (VEV) profile of the form $\langle H(z)\rangle={\rm diag}(v_5,v_5)/\sqrt{2}$ where $v_5(z)\equiv vR'/R^{3/2}
\sqrt{2(1+\beta)}(z/R')^{2+\beta}$, with $v\simeq 256\,$GeV and $\beta=0$, as arises in gauge-Higgs unification models where the Higgs is realized as a PGB~\cite{GHU}. In the color sector we extend the QCD gauge symmetry to SU(3)$_L\times$SU(3)$_R$.  
This extended symmetry is broken down to SU(3)$_V$ in the UV by boundary conditions and on the IR brane by the VEV of an EW singlet scalar field $\phi$ transforming as $({\bf 3,\bar3})$, allowing for SM quark masses to arise.\footnote{We assume that the colored PGB $\propto\arg(\phi)$ are above the few TeV range (see {\it e.g.} Ref.~\cite{DaRold:2005zs} for a related discussion in holographic QCD) and therefore leave no imprints in the current collider data. Although a lighter (possibly sub-TeV scale) colored PGB can be obtained, the associated collider phenomenology is outside the scope of this paper and we leave a dedicated analysis for future works.}
Such a ``custodial-like'' color bulk symmetry was already considered in Ref.\cite{Bauer:2011ah} as a attempt to solve the CP problem of anarchic RS models. We analyse in Appendix~\ref{App_kaon} the CPV contributions to both $K-\bar K$ mixing and radiative $K$ decays arising in SU(3)$_L\times$SU(3)$_R$ models with flavor anarchy and show that only a minor improvement relative to original anarchic RS models can be obtained (yet at the prize of a significantly worse Higgs mass fine-tuning, as shown in Appendix~\ref{finetuning}), so that a CP problem still remains. We resort instead to alignment in the down sector in order to evade the CPV strong constraints in the Kaon system.
The SU(3)$_L\times$SU(3)$_R$
bulk symmetry is introduced here in order to induce a large top $A_{\rm FB}$, similarly to axigluon models in 4D~\cite{Axigluons}.
The gauge field of the unbroken SU(3)$_V$ symmetry in 4D is identified to the SM massless gluon. In addition there are two towers of massive color octet KK states. One finds one tower of ``vector'' KK gluons associated with the 5D SU(3)$_V$ symmetry and another tower of ``axial'' KK gluons related to the coset SU(3)$_L\times$SU(3)$_R/$SU(3)$_V$. We denote the lightest vector (axial) KK resonance and its mass as $G^1_\mu$  ($A_\mu^1$) and $m_V$ ($m_A$) respectively. The $m_V$ is commonly referred to as the KK scale. Note that the axial states are generically heavier than the vector ones ($m_A\gtrsim m_V$) since the former receive additional contributions to their mass from the breaking of the axial symmetry. We define $t\equiv g_L/g_R$ to be the ratio between the SU(3)$_L$ and SU(3)$_R$ 5D gauge couplings. The SM chiral fermions are embedded into 5D Dirac fermions. We embed the quarks of the first and second generations in
\beq\label{embedding12}
Q^i_{U,D}\sim({\bf 3,1,2,\bar2})_{2/3,-1/3} \,,\ 
U^i\sim({\bf 1,\bar3,1,1})_{2/3} \,,\ D^i\sim({\bf 1,\bar3,1,1})_{-1/3} \,,
\eeq
where  $i=1,2$ is a generation index and $(L_3,R_3,L_2,R_2)_X$ denote the representations under SU(3)$_L\times$SU(3)$_R\times$SU(2)$_L\times$SU(2)$_R\times$U(1)$_X$. For the third generation we ``twist'' the representations in the color sector relative to the first two generations,
\beq\label{embedding3}
Q^3_{U,D}\sim({\bf 1,\bar3,2,\bar2})_{2/3,-1/3} \,,\ 
 U^3\sim({\bf 3,1,1,1})_{2/3} \,,\ D^3\sim({\bf 3,1,1,1})_{-1/3} \,,
\eeq
As we argue in Section~\ref{AFB} the above choice of representation allows for a positive top $A_{\rm FB}$ contribution from the axial KK gluon~\cite{Axigluons}. We assume the following representations for the leptons
\beq
L^i\sim ({\bf 1,1,2,\bar2})_{-1}\,, \ E^i\sim({\bf 1,1,1,1})_{-1}\,,
\eeq
where here the index $i=1,2,3$ runs over the three generations. 
Note that  we have embedded all 3 generations of RH SM fermions in singlet representations of SU(2)$_L\times$SU(2)$_R$ in order to avoid overly large non-oblique corrections to the EW parameters~\cite{FTRS,RedWei}. This choice, in turn, requires to embed each LH doublet of SM quarks into two 5D fields $Q_{U,D}$  bi-doublet of SU(2)$_L\times$SU(2)$_R$~\cite{AgasheDarold}. We assume appropriate boundary conditions of the 5D fermion fields (see Appendix~\ref{BCpsi}) such that the zero-modes are identified with the SM chiral fermions. In the basis where bulk fermion masses are diagonal in flavor space the couplings of $G_\mu^1$ and $A_\mu^1$ to SM chiral quarks are (See Appendix~\ref{KKdefs})
\beq\label{KKcpls}
g_{f}\left[\bar{f}\gamma^\mu  G_\mu^1f+s(t)\bar{f}\gamma^\mu A_\mu^1f\right] \,,\quad g_f\simeq g_V [\chi_f^2\gamma_f-\xi^{-1}]\,, 
\eeq
where $\xi=\log(R'/R)$, $g_V$ is the QCD KK coupling, $\gamma_f\simeq 2.3/(3-2c_f)$ and
$
\chi_f^2=(1-2c_f)/( 1-e^{-\xi(
1-2c_f)})\,
$
with $c_f$ the eigenvalues of the fermion bulk mass matrices (in units of $k$) defined as 
\beq
\mathcal{L}_{C}= \bar Q_U C_{Q_U} Q_U + \bar Q_D C_{Q_D} Q_D + \bar U C_{U}U + \bar D C_{D} D\,.
\eeq
We use the convention in which $c>1/2$ $(c<1/2)$ corresponds to elementary (composite) fermions.
$s(t)=-t$ for SM quarks embedded in $({\bf 3,1})$ of SU(3)$_L\times$SU(3)$_R$ and $s(t)=t^{-1}$ for SM quarks which belong to a $({\bf 1,3})$.

We will assume in the remainder of the paper the following set of parameters, which  illustrates the flavor paradigm advocated above
\beq
C_{Q_U}=C_{Q_D}\simeq (0.49,0.49,0.10) \,,\ C_U\simeq (0.41,0.40,0.48) \,,\ C_D\simeq (0.42,0.11,0.11)\label{bulkC}\,,
\eeq
where for simplicity we chose $C_{Q_U}=C_{Q_D}\equiv C_Q$, together with 
\beq\label{benchmark}
m_A\simeq 1.05 m_V\simeq 2.1\,{\rm TeV}\,,\ g_V\simeq 4\,,\ t\simeq 0.4\,.
\eeq
We motivate the above choice of flavor parameters both from EWPTs and flavor physics. We will discuss in details flavor physics considerations in Section~\ref{flavor}. We consider now EWPTs. A non-zero $T$ parameter arises at loop level despite the custodial symmetry of the EW gauge sector. At one-loop the dominant contribution is controlled by the large  5D top Yukawa. As a result of the left-right symmetric embedding of the third generation quarks under the custodial symmetry the loop induced $T$ parameter is typically negative~\cite{CarPon2006}. Together with a generically positive $S$ parameter a negative $T$ parameter pushes the KK scale to large values, thus re-introducing an unacceptable amount of fine-tuning of the weak scale. A way out is to make the right-handed (RH) top quark more elementary ($c_{U^3}\gtrsim 0.48$), which triggers a positive $T$ parameter at one-loop~\cite{CarPon2006}. It was shown in Ref.~\cite{FTRS} that in such a case the KK scale can be lowered down to $\mathcal{O}(2\,$TeV$)$ provided the light SM fermions have a rather flat profile in the bulk. The parameters' values in Eqs.~\eqref{bulkC}-\eqref{benchmark} are a slight deformation of the setup of Ref.~\cite{FTRS} 
and we have checked that the model outlined above is consistent with EW precision observables at the 95$\%$ confidence level.

Moreover the above setup constitutes an illustration of flavor models of class (IV), which displays exhilarated flavor structures. Indeed, as we argue in Section~\ref{flavor}, most severe flavor constraints arising from FCNC processes are satisfied thanks to alignment in the RH down and LH up sectors, while the observed asymmetries in the up sector (top  $A_{\rm FB}$ and charm $\Delta a_{CP}$) are reproduced within one standard deviation through composite light RH quark flavors.

Note also that for the above parameters  all couplings in Eqs.~\eqref{KKcpls} are individually perturbative. However, the width of the $A_\mu^1$ state is very broad $\Gamma_A^1\simeq 0.7m_A$. This signals that the above model, yet still perturbative, is close to a non-perturbative regime. Finally note that we have also assumed a quite degenerate KK-gluon spectrum ($m_A\simeq m_V$) in order to enhance the NP contributions to the top asymmetry. We justify this choice in Appendix~\ref{splitting}.

\section{Top forward-backward asymmetry}\label{AFB}
Both Tevatron experiments reported large $A_{\rm FB}$ measurements in various channels, all of them in some tension with the SM predictions. In contrast a very good agreement with QCD was found for the inclusive and differential $t\bar t$ cross-section. We focus below on the inclusive asymmetry $A_{\rm FB}^i$ as well as its value in the high invariant mass region ($m_{t\bar t}>450\,$GeV) $A_{\rm FB}^{450}$, for which the deviations from the SM expectations are most pronounced. CDF  measured the inclusive asymmetry both in the semi-leptonic~\cite{AFBCDF} and dilepton~\cite{AFBCDFdiL} $t \bar t$ 
channels, while D0  reported it only in the semi-leptonic one~\cite{AFBD0}. Since these 3 measurements are consistent with each other we chose to combine them through a weighted average, which yields 
\beq\label{AFBiexp}
A_{\rm FB}^i  \simeq 18\% \pm 4\% \, ({\rm
stat.+syst.}) \,,
\eeq
where the statistical and systematic errors were added in quadrature. Moreover CDF measured a significant growth of the asymmetry with $m_{t\bar t}$~\cite{AFBCDF}. The same trend exists in the D0 data although it is not very significant. Recall that the D0 measurement of $A_{\rm FB}^{450}$  is not corrected for detector's acceptance and efficiency and such effects can largely affect the result. For instance, the high mass CDF value at the partonic level is about twice as large as the background subtracted one before unfolding. We point out that applying an approximate unfolding factor of $\sim 2$ to the production level  high $m_{t\bar t}$ $A_{\rm FB}$ measurement at D0 shows better agreement with the CDF result. Thus, we chose to combine this naively unfolded D0 data with the parton level CDF result through a weighted average. We find 
\beq\label{AFB450exp}
A_{\rm FB}^{450}  \simeq 28\% \pm 6\% \, ({\rm
stat.+syst.}) \,.
\eeq
For the above two observables we consider the state of the art QCD+EW predictions as used in Ref.~\cite{AFBCDF}, which are 
\beq \label{AFBSM}
A_{\rm FB}^{i,{\rm SM}} = 6.6\%\,,\ A_{\rm FB}^{450,{\rm SM}} = 10\%\,.
\eeq
It is worth recalling that the above SM values are obtained using the next-to-leading order (NLO) total cross-section when computing the SM $A_{\rm FB}$, while a fixed order NLO calculation would require to use the leading order (LO) one. This results in an underestimate of  $\sim\mathcal{O}(30\%)$ of the SM asymmetries. Such a difference can be interpreted as the level of sensitivity of the result to the (yet unknown) higher order corrections. Comparing Eq.~\eqref{AFBiexp} and Eq.~\eqref{AFB450exp} to Eq.~\eqref{AFBSM}, NP contributions are required at the level of
$\delta A_{\rm FB}^i\simeq 11 (6)\%$ and $\delta A_{\rm FB}^{450}\simeq 18 (11)\%$. The numbers in parenthesis are the NP contributions needed to reach the measured values down by one standard deviation adding the experimental and theoretical errors in quadrature. 
The good agreement of the measured cross-section with the SM expectation in turns strongly constrains NP sources for $A_{\rm FB}$. Since our model involves NP states whose masses are above Tevatron's energies it is mostly constrained by the differential cross-section measurement in the hard region. We choose to represent the latter by the following large $m_{t\bar t}$ bin
$\sigma_{t\bar t}^h \equiv \sigma_{t\bar t}(700\,{\rm GeV}<m_{t\bar t}<800\,{\rm GeV})$.
The corresponding CDF measurement~\cite{CDFdiffXS}
\beq\label{XSh}
\sigma_{t\bar t}^{h,{\rm CDF}}= 80\pm 37\,{\rm fb}
\eeq
is very well consitent with the SM prediction~\cite{SMdiffXS}  
$\sigma_{t\bar t}^{h,{\rm CDF}}= 80\pm 8\,{\rm fb}$. Therefore we require the NP to SM cross-section ratio for this bin, $R_h$, not to exceed $\simeq 50\%$ in absolute value in order to remain within one standard deviation of Eq.~\eqref{XSh}.

We consider now the contributions of the first color-octet vector and axial KK states on $t\bar{t}$ $A_{\rm FB}$ and invariant mass distribution  at the Tevatron. We write the cross-section for top pair production as 
\beq
\sigma_{t\bar t}= \sigma_{t\bar t}^{\rm SM}(1+R)\,,\  {\rm where}\ 
R\equiv \frac{\sigma_{t\bar t}^{\rm NP}}{\sigma_{t\bar t}^{\rm SM}}
\eeq
is the ratio of NP to SM the cross-sections. The asymmetry reads
$A_{\rm FB}= A_{\rm FB}^{\rm SM}+\delta A_{\rm FB}$, where $A_{\rm FB}^{\rm SM}$ is the SM prediction and $\delta A_{\rm FB}$ denotes the NP contribution which we write as  
\beq\label{dAFB}
\delta A_{\rm FB}\equiv \frac{A_{\rm FB}^{\rm NP}}{(1+R)}-\frac{A_{\rm FB}^{\rm SM}}{(1+1/R)}\,,
\eeq
where we defined $A_{\rm FB}^{\rm NP}$ as the ratio of the NP asymmetric cross-section to the SM symmetric one; The second term in Eq.~\eqref{dAFB} arises from a deviation in the symmetric cross-section relative to the SM. 
We chose to express the observables in terms of NP to SM ratios of cross-section in order to minimize the impact of the unknown NLO corrections to the NP contributions. We argue that the NP to SM ratios evaluated at LO would constitute a good approximation to equivalent ratios at NLO since the Lorentz and gauge structures of the NP is similar to that of QCD. We  thus compute $R$ and $A_{\rm FB}^{\rm NP}$ at LO and we use the most accurate prediction available for the SM expectations $A_{\rm FB}^{\rm SM}$ and $\sigma_{t\bar t}^{\rm SM}$. 

\begin{figure}[tb]
\begin{center}
\includegraphics[width=0.6\textwidth]{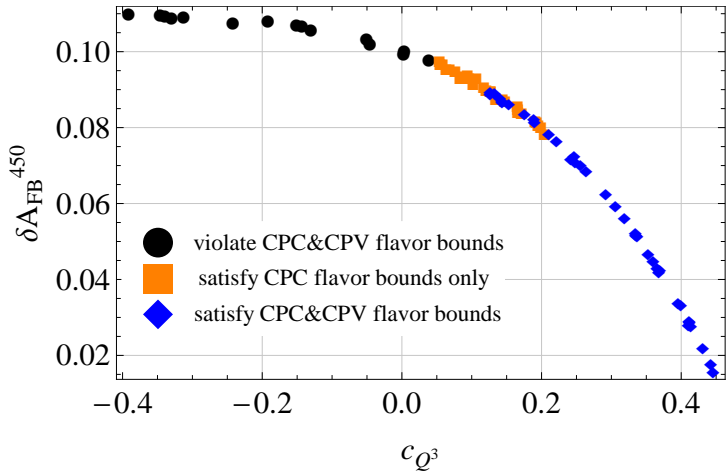}
\end{center}
\caption{$\delta A_{\rm FB}^{450}$ as function of $c_{Q^3}$ assuming other parameters fixed to the values in Eqs.~\eqref{bulkC}-\eqref{benchmark}. The points correspond to different values of $c_{Q^1}$ and $c_{U^3}$ which we randomly varied  in $c_{Q^1}\in [0.47,0.53]$, and $c_{U^3}\in [0.4,0.49]$. All points are consistent with EWPTs. Orange points satisfy all CP concerving (CPC) flavor violation bounds but violate at least one CPV flavor bound. Blue points satisfy all CPC and CPV flavor violation bounds, whereas black point satisfy neither of them.} 
\label{fig_flavor-afb}
\end{figure}

We proceed as follows to derive the NP contributions to the above top observables arising from the model defined in Section~\ref{section_model}. Since the relevant KK resonances have masses of $\mathcal{O}(2\,{\rm TeV})$ their effects on $t\bar t$ observables at Tevatron are well described in terms of the following effective Lagrangian\footnote{We choose to work in a vector/axial basis for convenience. See also Refs.~\cite{Ko,Willenbrock,Maltoni} for similar effective field theory analysis in an SU(2)$_L$ invariant basis.} 
\beq\label{EFTtop}
\mathcal{L}_{\rm eff}^{\bar t t}= \frac{c^8_{V}}{\Lambda^2}\mathcal{O}_{V}^8+\frac{c^8_{A}}{\Lambda^2}\mathcal{O}_{A}^8+\frac{c^8_{VA}}{\Lambda^2}\mathcal{O}_{VA}^8+\frac{c^8_{AV}}{\Lambda^2}\mathcal{O}_{AV}^8
\eeq
where 
\bea
\mathcal{O}_{V}^8 &=& (\bar{u}\gamma_\mu T^a u)(\bar{t}\gamma^\mu T^a t)\,,\label{OV8}\\ 
\mathcal{O}_{A}^8 &=& (\bar{u}\gamma_\mu\gamma_5 T^a u)(\bar{t}\gamma^\mu\gamma_5 T^a t)\,,\label{OA8}\\
\mathcal{O}_{VA}^8 &=& (\bar{u}\gamma_\mu T^a u)(\bar{t}\gamma^\mu\gamma_5 T^a t)\,,\label{OVA8}\\ 
\mathcal{O}_{AV}^8 &=& (\bar{u}\gamma_\mu\gamma_5 T^a u)(\bar{t}\gamma^\mu T^a t)\,.\label{OAV8}
\eea
We use the 
expressions in Ref.~\cite{WISEFT} to relate $A_{\rm FB}^{i,{\rm NP}}$, $A_{\rm FB}^{450,{\rm NP}}$ 
and $R_{i,450,h}$ to the Wilson coefficients of the operators listed in Eqs.~\eqref{OV8}-\eqref{OAV8}. $R_{i}$ and $R_{450}$ are the NP to SM cross-section ratios at the inclusive level and for large invariant masses satisfying $m_{t\bar t}>450\,$GeV, respectively.
Note that only $\mathcal{O}_{V,A}^8$ interfere with the parity even leading order QCD amplitude and their effects are thus expected to dominate. We focus only on these two interfering operators to derive some simple relations valid at $\mathcal{O}(\alpha_s/\Lambda^2)$; our numerical results include all effects up to $\mathcal{O}(1/\Lambda^4)$. $\mathcal{O}_V^8$ contributes to the symmetric cross-section and $\mathcal{O}_A^8$ only contributes to the asymmetry. We find the following expressions for their corresponding Wilson coefficients
\beq
C_{V,A}^8\equiv \frac{c^8_{V,A}}{\Lambda^2}= -\frac{\left(g_{Q^1}\pm g_{U^1}\right)\left(g_{Q^3}\pm g_{U^3}\right)}{4m_V^2}+\frac{\left(g_{Q^1}t\mp g_{U^1}t^{-1}\right)\left(g_{Q^3}t^{-1}\mp g_{U^3}t\right)}{4m_A^2}\,.
\eeq
A positive $C_A^8$ yields a positive $A_{\rm FB}$ contribution at the interference level~\cite{WISEFT}.
Note that in the absence of the axial KK gluon ($m_A\gg m_V$) the asymmetry contribution is positive and maximal when the KK gluon couples only to up and top of opposite chiralities, $u_L$ and $t_R$ or $u_R$ and $t_L$. In both cases $C_A^8\simeq -C_V^8$ and the KK gluon distords the symmetric cross-section as much as it contributes to $A_{\rm FB}$. At $\mathcal{O}(\alpha_s/\Lambda^2)$ we find approximately~\cite{WISEFT}
\beq
\delta A_{\rm FB}^{450}\simeq -0.2 R_h\,,
\eeq  
where $R_h<0$ for $C_A^8>0$. Hence, requiring that the deviation relative to the SM for the high $m_{t\bar t}$ bin is less than $50\%$ results in an upper bound on the asymmetry of $\delta A_{\rm FB}^{450}\lesssim 10\%$. Although this is the right order of magnitude to explain the data, this value is only obtained through an overly large up quark compositeness conflicting with dijet seaches. This is understood as follows. $\delta A_{\rm FB}^{450}\sim 10\%$ requires $C_A^8\sim 0.5/$TeV$^2$~\cite{WISEFT}. In order to minimize the dijet contribution 
 we assume $C_A^8 \simeq g_{U^1}g_{Q^3}/4m_V^2$ so that NP dijet events are induced by only one light quark flavor (RH up). 
Under the small dijet contribution requirement, the large $A_{\rm FB}$ implies a maximal top coupling and a low KK scale. EWPTs allow $m_V\simeq 2\,$TeV with $c_{Q^3}\simeq 0$~\cite{FTRS}. This yields $g_{Q^3}\lesssim 5$, where we took the 5D gauge coupling to saturate its perturbative value $g_V\lesssim 4\pi/\sqrt{3}\sim 7$, and $g_{U^1}\gtrsim 1.6$. 
This implies a rather low compositeness scale for the RH up quark
, which is a factor of $\sim2$ below the bound imposed by present experimental constraints~\cite{Pomarol}. Dijet searches therefore impose $g_{U^1}\lesssim 1$ and so $\delta A_{\rm FB}^{450}\lesssim 5\%$. This demonstrates the need to introduce another source of $A_{\rm FB}$ in warped/composite models. We show in Fig.~\ref{fig_flavor-afb}  that a sufficiently large asymmetry is induced in complete agreement with EWPT, flavor and dijet constraints for the model defined in Section~\ref{section_model}. The various points in Fig.~\ref{fig_flavor-afb} are obtained by varying randomly $c_{Q^1}$ and $c_{U^3}$ around the values in Eq.~\eqref{bulkC}, while other flavor parameters are adjusted to reproduce the CKM mixing angles as well as the SM quark masses. In particular, we see that $c_{Q^3}<0$, which leads to $\delta A_{\rm FB}^{450}\gtrsim10\%$, is excluded by flavor physics. The maximal $A_{\rm FB}$ contribution above $m_{t\bar t}>450\,$GeV is obtained for the parameters in Eqs.~\eqref{bulkC}-\eqref{benchmark}, which yield $\delta A_{\rm FB}^{450}\simeq 10\%$. For the same parameters, we also find $\delta A_{\rm FB}^i\simeq 4\%$, which is about $1.3\sigma$ below the measurement, and a small contribution to the $t\bar t$ distribution at Tevatron of $|R_h|\simeq 25\%$. We also find that the charge asymmetry at the 7 TeV LHC is consistent with the ATLAS~\cite{ATLASac} and CMS~\cite{CMSac} measurements both at the inclusive and differential levels.\\

\section{CP violating charm decays}\label{section_CPV-LHCb}
Recently the LHCb collaboration reported  the first evidence~\cite{lhcb} for a non-zero
value of the difference between the time-integrated CP asymmetries in the
decays $D^0 \to K^+K^-$ and $D^0 \to \pi^+\pi^-$, respectively denoted by $a_{CP}^{KK}$ and $a_{CP}^{\pi\pi}$. Combined with other measurements of these CP asymmetries~\cite{BelleICHEP,CDF10784, Aaltonen:2011se, Staric:2008rx, Aubert:2007if,HFAG}, the present world average is 
\begin{equation}
\Delta a_{CP} \equiv a_{CP}^{KK} - a_{CP}^{\pi\pi} = -(0.74\pm 0.15)\%\,,
\label{eq:acpExp}
\end{equation}
which is $\simeq 4.9\sigma$ away from zero. We assume that this result is dominated by NP.
We argue that this new CP violating effect may have interesting implications for new physics searches at the LHC. In particular we demonstrate that in models where $\Delta a_{CP}$ is accomodated by $s$-channel FCNCs the strong constraint from $D-\bar D$ mixing combined with constraints from dijets searches at the LHC implies that  the NP source of CPV is only at play in the $K^+K^-$ channel. 

The NP contribution to $\Delta a_{\rm CP}$ are fully parameterized by a complete set of $\Delta C=1$ effective operators at the charm scale. As shown in Ref.~\cite{IsiKam} only a few operators accomodate the LHCb result without conflicting with present bounds from either $D-\bar{D}$ mixing or $\epsilon'/\epsilon_K$. A particularly interesting way to induce direct CPV in charm decay is through a chromomagnetic dipole operator, a possibility already broadly discussed in a supersymmetric context in Refs.~\cite{NirKagan,SUSYdipoles} and also in Refs.~\cite{Rattazzi,DelKam} in warped/composite models.
We focus here on another attractive NP path towards charm CPV which relies on the following part of the $\Delta C=1$ effective Lagrangian
\beq\label{LeffaCP}
\mathcal{L}_{\rm eff}^{\Delta C=1}\supset \sum_{q=u,d,s}\frac{1}{\Lambda_{a_{CP}}^{q\,2}}\mathcal{O}_{\bar{u}c \bar q q}^8+{\rm h.c.}\,\quad {\rm where}\quad
\mathcal{O}^8_{\bar{u}c\bar q q}=(\bar{u}_R\gamma^\mu T^a c_R)(\bar{q}_R\gamma_\mu T^a q_R). 
\eeq
There is another independent color contraction for each operator above. 
Since we match both the axial and vector KK gluons to the Lagrangian Eq.~\eqref{LeffaCP} we assume implicitly below a color octet contraction. 
Note however that the nuclear matrix elements for the $\Delta C=1$ operators in Eq.~\eqref{LeffaCP} are unknown such that it is not possible to differentiate the effect of a singlet contraction from that of a octet one. 
Given that the observation of CPV is only significant in the difference of CP asymmetries any of the operators in Eq.~\eqref{LeffaCP} can generate the $\Delta a_{CP}$ measurement. Because only RH quark currents are involved the above operators do not lead to overly large flavor violation in the down-type quark sector and are thus safe from the strong $\epsilon'/\epsilon_K$ constraint~\cite{IsiKam}. The experimental result in Eq.~\eqref{eq:acpExp} suggests a scale of~\cite{IsiKam}
\beq\label{DaCPscale}
\Lambda_{a_{CP}}^q\simeq 13\,{\rm TeV}  
\eeq
for any $q=u,d,s$. The scale above comes with an $\mathcal{O}(1)$ uncertainty due to the unknown hadronic matrix elements relevant to the $D^0\to \pi^+\pi^-$ and $D^0\to K^+K^-$ decays~\cite{IsiKam}.

We now show that in NP models where $\mathcal{O}_{\bar u  c \bar q q}^8$ arises from a heavy state exchanged in the $s$-channel the $\Delta a_{CP}$ measurement has an immediate implication for flavor and CP conserving observables at the TeV scale. Indeed, in this case, the same NP producing the Lagrangian Eq.~\eqref{LeffaCP} will also give rise to the following $\Delta C=2$ Lagrangian
\beq
\mathcal{L}_{\rm eff}^{\Delta C=2} \supset \frac{1}{\Lambda_{\bar{u}c}^2}\mathcal{ O}_{\bar u c}^8+{\rm h.c.}\quad {\rm with}\quad \mathcal{O}_{\bar{u} c}^8=\frac{1}{2}(\bar u_R \gamma_\mu T^a c_R)^2\,
\eeq
as well as the flavor conserving ``compositeness'' four-fermion interactions 
\beq
\mathcal{L}_{\rm eff}^{\Delta C=0} \supset \sum_{q = u,d,s}\frac{1}{\Lambda_{q q}^2}\mathcal{O}_{q q}^8\quad {\rm with}\quad \mathcal{O}_{ q q}^8=\frac{1}{2}(\bar q_R \gamma_\mu T^a q_R)^2
\eeq
where the following relation among the corresponding operator scales is predicted
\beq\label{Dseesaw}
\Lambda_{a_{CP}}^q\simeq
\sqrt{\Lambda_{q q}\Lambda_{\bar u c}}\,.
\eeq
The operator ${\mathcal O}_{\bar u c}^8=\frac{1}{3}{\mathcal O}_{\bar u c}^1$, where $\mathcal{O}_{\bar u c}^1\equiv \frac{1}{2}(\bar u_R \gamma^\mu c_R)^2$, contributes to $D-\bar{D}$ mixing. Assuming a maximal CPV phase the corresponding scale is therefore severely constrained. The present bound on $\mathcal{O}_{\bar u c}^1$ (see {\it e.g.} Ref.~\cite{INP}) implies
\beq\label{DDbarbound}
\Lambda_{\bar u c}\gtrsim 490\,{\rm TeV}\,.
\eeq  
We used the bound on the real part of the Wilson coefficient of $\mathcal{O}_{\bar u c}^1$, although the bound on the CP violating part is stronger. This choice is conservative in the sense that the bound on the imaginary part may be relaxed if indeed there is an active NP source of CPV at play in the $D$ system, as the $\Delta a_{CP}$ measurement suggests.
The flavor conserving operator $\mathcal{O}_{q q}$ characterizes the degree of compositeness of the light quark flavors.
Using Eqs.~\eqref{DaCPscale} and~\eqref{DDbarbound} the relation Eq.~\eqref{Dseesaw} implies 
\beq
\Lambda_{\bar q q}\lesssim 350\,{\rm GeV}\,.
\eeq
The widly separated scales of $\Delta C=2$ and $\Delta C=1$ processes, given in Eq.~\eqref{DDbarbound} and Eq.~\eqref{DaCPscale} respectively, thus generically requires a very low compositeness scale for the light quark flavors. We call this effect the new physics ``charm seesaw''.
For the valence quarks $q=u,d$ a compositeness scale of $\sim 300\,$GeV is already excluded by dijet searches at the Tevatron~\cite{CDFdijets,D0dijets}. Quark compositeness is also probed through dijet searches at the LHC~\cite{CMSdijets,ATLASdijets}. Present dijet searches at ATLAS and CMS are consistent with SM predictions and the $u_R$ and $d_R$ compositeness scale is constrained to be $\Lambda_{u u}\gtrsim 2.1\,$TeV and $\Lambda_{d d}\gtrsim 1.5\,$TeV, respectively~\cite{Pomarol}. On the other hand dijet searches are less sensitive to contact interactions involving the strange quark since the latter has a suppressed parton distribution function (PDF) inside a proton. We show in Section~\ref{sec:dijets} that dijets production from the operator $\mathcal{O}_{ s s}$ at a scale $\Lambda_{s s}\sim 300\,$GeV is consistent with  the current LHC data. We conclude that an $\mathcal{O}^8_{\bar u c\bar s s}$ operator induced by a $s$-channel exchanged heavy octet can accomodate the $\Delta a_{\rm CP}$ measurement without conflicting with present constraints from $D-\bar D$ mixing, $\epsilon'/\epsilon_K$ and dijet searches.

The NP scenario for $\Delta a_{CP}$ outlined above has several generic predictions both for charm and high $p_T$ physics. All of them can be tested in the near  futur. First of all, since only the strange quark is very composite CPV is predicted to be mostly at work in the $D^0\to K^+K^-$ channel, a prediction soon to be tested by flavor experiments by measuring the CP asymmetries for each channel individually. To the best of our knowledge this prediction is unique and differs from all other proposed explanation of $\Delta a_{CP}$ based on either SM-like CPV sources~\cite{DaCPSM} or NP contributions from CPV chromomagnetic dipoles~\cite{NirKagan,SUSYdipoles,Rattazzi,DelKam}.
Second, assuming a maximal NP phase CPV is expected to be observed in $D-\bar D$ mixing in the near future. Finally the very composite strange quark required by this scenario would leave an imprint in the angular distribution of dijets produced at the LHC, a signature which should be visible in the 8 TeV run. 

The composite models analysed in this paper are of the type outlined above. The situation is however slightly different when the axial KK gluon is included in the spectrum since in this case the Lagrangian Eq.~\eqref{LeffaCP} receives contributions from two resonances, the KK vector and axial gluons. As described in Section~\ref{flavor}, the two resonances have different flavor violating couplings to quarks due to the different SU$(3)_L\times$SU$(3)_R$ embedding of the third generation relative to the the first two.
Using the flavor ansatz oulined in Section~\ref{flavor} we have checked that the scale of CPV in charm decay can be as low as $\Lambda^s_{a_{CP}}\simeq 26\,$TeV in exhilarated flavor RS models, while all flavor constraints from $\Delta F=2$ processes are satisfied and $A_{FB}^{450}\sim10\%$. The above scale is larger than the scale required in Eq.~\eqref{DaCPscale}. However, given the large uncertainties in the relevant hadronic matrix element, the NP contribution to CPV in charm decays can be sufficiently large to accommodate the $\Delta a_{CP}$ measurement. Furthermore the new CPV phase inducing $\Delta a_{CP}$ is also at play in $D-\bar D$ mixing. As argued above the bound on CP conserving $D-\bar D$ mixing is typically saturated in order to reduce the RH strange compositeness. Therefore, under the assumption that the new phase is maximal, our model predicts that CPV in $D-\bar D$ mixing should be observed in the near future. As a result the bound on CPV in $D-\bar D$ mixing would be relaxed. Under all the above requirements but removing constraint from CP violating $D-\bar D$ mixing (see Table~\ref{tableCij}), we find that a minimal scale of $\Lambda^s_{a_{CP}}\simeq 19\,$TeV in exhilarated flavor RS models, which yields a $\Delta a_{CP}$ value within one standard deviation.

\section{Flavor physics constraints}\label{flavor}

The huge improvement over the recent years in the determination of the CKM matrix elements and in the constraints on new physics sources of flavor changing neutral current (FCNC) processes strongly restrict the flavor structure of NP at the TeV scale. Tree level exchange of KK states between partially composite light quarks can mediate dangerous $\Delta F=2$ processes in the form of meson-antimeson mixing. These effects are generally described using an EFT expansion at the appropriate meson scale. The effective Lagrangian relevant to our analysis is
\beq
\mathcal{L}_{\rm eff}^{\Delta F=2}\supset \sum_i\frac{c_i}{\Lambda^2}\mathcal{O}_i+\frac{c_i'}{\Lambda^2}\mathcal{O}_i'
\eeq
where the sum runs over the following operators
\bea
\co_1&=&\bar q^{j\alpha}_L\gamma_\mu q^{i\alpha}_L\bar q^{j\beta}_L\gamma_\mu q^{i\beta}_L\,,\\
\co_4&=&\bar q^{j\alpha}_Rq^{i\alpha}_L\bar q^{j\beta}_Lq^{i\beta}_R\,,
\eea
where $q^i$ denotes the SM quark field of flavor $i$ (with $i\neq j$) and $\alpha,\beta$ are color indices. Primed operators are obtained by the replacement  $L\to R$. We match the Wilson coefficients $C^i\equiv c_i/\Lambda^2$ and $C^{i\prime}\equiv c'_i/\Lambda^2$ to the vector and axial KK gluon resonances and we determine the quark compositeness and the Yukawa coupling structure which satisfy both flavor and dijet constraints and address the flavor anomalies considered in the previous sections.

Although we set by hand the flavor parameters of the model in order to reproduce the experimental results, we argue that the data points towards a certain flavor structure. To a good approximation this structure consists in a 4D alignment with SM mass matrices of the right-handed (RH) down-quark and left-handed  (LH) up-quark sectors, whereas the RH up-quark and LH down-quark sectors are misaligned. We show below how such a flavor structure is consistent with flavor (and CP) violating observables.

We define the flavor basis as the basis where the fermion bulk masses are diagonal in flavor space. The mass matrices of the SM-like zero modes in the flavor basis are
\begin{equation}\label{mfl}
M_U\propto F_{Q_U}^\dagger Y_U F_U\,,\quad M_D\propto F_{Q_D}^\dagger Y_D F_D\,,
\end{equation}
where $F_{X}={\rm diag}(\chi_{X^1},\chi_{X^2},\chi_{X^3})$ with $X=Q_{U},Q_D,U,D$. The basis in which $M_{U,D}$ are diagonal is referred to as the mass basis. The flavor and mass basis are connected by a set of unitary transformations $V_X$ of the zero-mode quark fields X, such that
\begin{equation}\label{mph}
{\rm diag}(m_u,m_c,m_t)=V_{Q_U}^\dagger M_U V_U\,, \quad {\rm diag}(m_d,m_s,m_b)=V_{Q_D}^\dagger M_D V_D\,.
\end{equation} 
The CKM matrix is $V_{CKM}=V_{Q_U}^\dagger V_{Q_D}$. The 4D alignment advocated above leads to $V_D\simeq{\bf 1}_{3\times 3}$ and $V_{Q_U}\simeq {\bf 1}_{3\times 3}$. 

Besides the fermion bulk masses defined in Eq.~\eqref{bulkC} we consider the following values for the 5D Yukawa coupling matrices defined in the flavor basis\footnote{The $\phi$ VEV insertion is understood wherever required by gauge invariance. We further assumed that $\phi$ is a localized field on the IR brane. We checked that taking $\phi$ to propagate in the bulk, while remaining peaked towards the IR brane, leads to similar quantitative results.}
\beq
\mathcal{L}_Y= \bar Q_U Y_U H U + \bar Q_D Y_D H D + {\rm h.c.}
\eeq
with appropriate gauge and flavor contractions and 
\bea
Y_U&=&\left(\begin{array}{ccc} -2.9\times 10^{-4}& 2.1\times 10^{-2} & -0.21 \\ 
-1.9\times 10^{-3}& 8.6\times 10^{-2} & -0.82 \\ 
1.2\times10^{-2} & 1.2\times10^{-2}& 4.2\end{array}\right)\,,\\ 
Y_D&=&\left(\begin{array}{ccc} 3.6\times10^{-2}& 0.21 & 0.10 \\
-1.8\times10^{-2} & 0.41 & 0.16 \\
-1.0\times10^{-5}  & -1.0\times 10^{-3} & 4.1\end{array}\right)\times10^{-2}\,.
\eea
We implicitly assume that all CP violating phases other than the KM phase are maximal. This defines completely the 5D flavor parameters of the model.
Using the usual parametrization for 3$\times$3 rotation matrices, the above flavor parameters lead to the following mixing angles $(\theta_{Q_U}^{12},\theta_{Q_U}^{23},\theta_{Q_U}^{31})\simeq(15^\circ,-2^\circ,-0.5^\circ)$ and $(\theta_{U}^{12},\theta_{U}^{23},\theta_{U}^{31})\simeq(0.29^\circ,0.3^\circ,0.3^\circ)$. 
We have checked that misalignment induced at leading order by $Y_D^\dagger Y_D$ and $Y_U Y_U^\dagger$ matrices are small, so that the flavor structure argued above is radiatively stable. We demonstrate below that this ansatz  satisfies the flavor constraints from $\Delta F=2$ processes and induces a large enough contribution to $\Delta a_{CP}$.

The vector and axial  KK gluon couplings  to quarks in the mass basis are needed in order to match the Wilson coefficients $C^i$ to the warped model defined above. Recall these couplings in the flavor basis are 
\beq
G_V^X={\rm diag}(g_{X^1},g_{X^2},g_{X^3})\,,\quad
G_A^X={\rm diag}(s_{X^1}(t)g_{X^1},s_{X^2}(t)g_{X^2},s_{X^3}(t)g_{X^3})\,,
\eeq
respectively, with $G_{V,A}^{Q_U}=G_{V,A}^{Q_D}$ thanks to SU(2)$_L$ symmetry. Notice that $s_{X^i}(t)$ is chirality and flavor dependent due to the embedding under the strong gauge symmetries defined in Eqs.~\eqref{embedding12} and~\eqref{embedding3}. 
The KK gluons couplings to quarks in the mass basis are
\begin{equation}
G_{V,A}^{X\,{\rm mass}}=V_X^\dagger G_{V,A}^X V_X \,,
\end{equation}
or, more explicitly,
\begin{eqnarray}
G_{V\,ij}^{X\,{\rm mass}}&=&V_X^{2i*}V_X^{2j} (g_{X^2}-g_{X^1})+V_X^{3i*}V_X^{3j} (g_{X^3}-g_{X^1})\,, \label{gijV} \\
G_{A\,ij}^{Q_{U,D}\,{\rm mass}}&=&-tV_{Q_{U,D}}^{2i*}V_{Q_{U,D}}^{2j} (g_{Q_{U,D}^2}-g_{Q_{U,D}^1})+V_{Q_{U,D}}^{3i*}V_{Q_{U,D}}^{3j} \left(t^{-1}g_{Q_{U,D}^3}+tg_{Q_{U,D}^1}\right) \,,\label{gijAL} \\
G_{A\,ij}^{U,D\,{\rm mas}}&=&t^{-1}V_{U,D}^{2i*}V_{U,D}^{2j} (g_{U^2,D^2}-g_{U^1,D^1})-V_{U,D}^{3i*}V_{U,D}^{3j} \left(tg_{U^3,D^3}+t^{-1}g_{U^1,D^1}\right) \,, \label{gijAR}
\end{eqnarray}
Notice that the axial KK gluon couplings are non-diagonal in flavor space, even for universal 5D masses. This is reminiscent of the twisting of the third generation under the strong gauge symmetries which breaks explicitly the flavor symmetries. This new source of flavor violation is absent for the vector KK gluon couplings, which are flavor conserving for universal 5D masses.
We summarize in Table~\ref{tableCij} the expressions for the matched Wilson coefficients and we derive the bounds on the appropriate combination of vector and axial flavor violating couplings using the results of Ref.~\cite{INP} and $g_V/m_V\simeq2/$TeV 
 following from Eq.~\eqref{benchmark}.
\begin{table}[tb]
\begin{center}
\renewcommand{\arraystretch}{2}
\begin{tabular}{c|c|cc}
\hline
\hline
Wilson coefficient & Matched value  & 
 \multicolumn{2}{c}{Bound on coupling} \\
 & (in units of [$g_V^2/m_V^2$])  &
 Re & Im \\
\hline
$C^1_K$ & 
$\frac{1}{6}\left[(G^{Q_D\,{\rm mass}}_{V\,21})^2+\delta^{-1} (G^{Q_D\,{\rm mass}}_{A\,21})^2\right]$ & 
$1.2\times 10^{-3}$ & $7.7\times 10^{-5}$ \\
$C^4_K$ & 
$G^{Q_D\,{\rm mass}}_{V\,21}G^{D\,{\rm mass}}_{V\,21}+\delta^{-1} G^{Q_D\,{\rm mass}}_{A\,21}G^{D\,{\rm mass}}_{A\,21}$ & 
$2.8\times 10^{-5}$ & $1.6\times 10^{-6}$ \\
\hline
$C^1_D$ & 
$\frac{1}{6}\left[(G^{Q_U\,{\rm mass}}_{V\,21})^2+\delta^{-1} (G^{Q_U\,{\rm mass}}_{A\,21})^2\right]$ & 
$1.0\times 10^{-3}$ & $4.2\times 10^{-4}$ \\
$C^4_D$ & 
$G^{Q_U\,{\rm mass}}_{V\,21}G^{U\,{\rm mass}}_{V\,21}+\delta^{-1} G^{Q_U\,{\rm mass}}_{A\,21}G^{U\,{\rm mass}}_{A\,21}$ & 
$8.1\times 10^{-5}$ & $3.3\times 10^{-5}$ \\
\hline
$C^1_{B_d}$ & 
$\frac{1}{6}\left[(G^{Q_D\,{\rm mass}}_{V\,31})^2+\delta^{-1} (G^{Q_D\,{\rm mass}}_{A\,31})^2\right]$ & 
$2.4\times 10^{-3}$ & $1.3\times 10^{-3}$ \\
$C^4_{B_d}$ & 
$G^{Q_D\,{\rm mass}}_{V\,31}G^{D\,{\rm mass}}_{V\,31}+\delta^{-1} G^{Q_D\,{\rm mass}}_{A\,31}G^{D\,{\rm mass}}_{A\,,31}$ & 
$2.6\times 10^{-4}$ & $1.4\times 10^{-4}$ \\
\hline
$C^1_{B_s}$ & 
$\frac{1}{6}\left[(G^{Q_D\,{\rm mass}}_{V\,32})^2+\delta^{-1} (G^{Q_D\,{\rm mass}}_{A\,32})^2\right]$ & 
\multicolumn{2}{c}{$1.1\times 10^{-2}$} \\
$C^4_{B_s}$ & 
$G^{Q_D\,{\rm mass}}_{V\,32}G^{D\,{\rm mass}}_{V\,32}+\delta^{-1} G^{Q_D\,{\rm mass}}_{A\,32}G^{D\,{\rm mass}}_{A\,32}$ & 
\multicolumn{2}{c}{$1.4\times 10^{-3}$} \\
\hline
\hline
\end{tabular}
\end{center}
\caption{Bounds on flavor violating couplings from $\Delta F=2$ processes. The first and second columns show the Wilson coefficient of the relevant four-fermion operators and the their matched values, respectively. 
The last column presents the bounds on the corresponding combination of flavor violating couplings (derived using the bounds presented in Ref.~\cite{INP}) assuming
$g_V/m_V=2/$TeV. 
The Wilson coefficients $C^{1\prime}$ are obtained from $C^1$ through the replacements $Q_U\to U$ and $Q_D\to D$.}
\label{tableCij}
\end{table}
We discuss flavor violation in the up-type and down-type sectors separately. 

\subsection{Flavor violation in the up-sector}
We consider here the constraints from $D$ meson mixing as well as the implications of the sizable CPV in charm decay induced in our model.
As argued in Section~\ref{section_CPV-LHCb}, a large NP source of CPV in charm decays is desired from the RH sector, without conflicting with the constraints from $D-\bar D$ mixing. Both effects are driven by the flavor violating couplings defined above which involve the up and charm quarks. Direct CPV requires a sizable coupling $G^{U\,{\rm mass}}_{V,A\, 21}$ in the RH sector, while flavor constraints in the $D$ system put severe bounds on the couplings $G^{U\,{\rm mass}}_{V,A\,21}$ and $G^{Q_U{\rm mass}}_{V,A\,21}$ in the RH and LH, respectively. In order to maximize direct CPV we assume $C^{1\prime}_D$ saturates the bound by taking $G^{U\,{\rm mass}}_{V,A\,21}\sim\co(10^{-3})$. This, in turn, requires $G^{Q_U\,{\rm mass}}_{V,A\,21}\lesssim \co(10^{-5})$ to comply with the bounds on $C^4_D$, while the bounds on $C^1_D$ are much less stringent.

We detail now the implications on other flavor violating observables of the above choice of couplings.  Equations.~\eqref{gijV}-\eqref{gijAR} show that there are different types of flavor violating sources present in the couplings with the vector and axial KK gluons. Those are flavor non-universality of the fermion bulk masses, misalignement between $F_{Q_U}$ ($F_{U}$) and $Y_UY_U^\dagger$ ($Y^\dagger_UY_U$) and the twisting of the third generation relative to the first two generations under SU(3)$_L\times$SU(3)$_R$. As argued in Section~\ref{AFB} a large top $A_{\rm FB}$ typically requires $t<1$ together with a sizable $Q_U^3$ compositeness. Thus the above upper bound on $G^{Q_U{\rm mass}}_{A\,21}$ limits the degree of misalignment in the LH up-quark sector.  From the second term of Eq.~(\ref{gijAL}), for $g_{Q_{U,D}^3}\sim \co(1)$, we obtain:
\beq\label{boundqu}
V_{Q_U}^{32*}V_{Q_U}^{31}\lesssim \co(10^{-6})\ .
\eeq
This estimate shows that, for our benchmark point, $D$ physics requires the mixings between the LH quarks of the first two generation and $t_L$ to be small. The first term of Eq.~(\ref{gijV}) also potentially induces overly large flavor violation. We choose to suppress these effects through degenerate bulk masses $c_{Q^1_U}=c_{Q^2_U}$.
We choose flavor non-universal $c_{u^i}$ in order to induce the mixing angle satisfying $G^{U\,{\rm mass}}_{V,A\,21}\sim\co(10^{-3})$.
It is straightforward to check that the ansatz stated above for the flavor parameters  is consistent with the flavor structure (LH up-quark alignment at leading order) advocated for the up-sector.

\subsection{Flavor violation in the down-sector}

We discuss here the constraints on flavor and CP violation in the $B$ and Kaon systems.
As argued in the previous subsection, constraints on $D-\bar D$ mixing require tiny mixing angle between the first two and the third generation in the LH up-sector, $V_{Q_U}\sim {\bf 1}_{3\times3}$. $V_{Q_D}$ must then be close to the CKM matrix, which could lead to large FCNCs in the down sector, especially when $s_R$ and $b_R$ are composite.
We first discuss constraints from Kaon physics.
We note that NP contributions to $K-\bar K$ mixing and Kaon decay from RH down currents are suppressed through alignment. However, the LH down mixing angles, being of CKM size, could lead to overly large contributions to $\epsilon_K$ and $\epsilon'/\epsilon_K$. We choose to suppress these NP contributions through degeneracy by taking $c_{Q_{U,D}^1}=c_{Q_{U,D}^2}$. There is therefore no tension with Kaon physics in our setup. As argued this is only possible thanks to the peculiar flavor structure of the down sector. This situation radically differs from the anarchic approach, which we study in detail in Appendix~\ref{App_kaon}.

We move now to discuss NP effects in $B$ physics.
Since $V_{Q_D}$ is close to the CKM matrix, dangerous corrections to the strongly constrained $B_{d}$ and $B_s$ mixings can be induced. Moreover, notice that the large degree of  $Q_U^3$ compositeness and $t<1$ required by a large top $A_{\rm FB}$, typically enhance NP contributions to $C^1_{B_d}$ and $C^1_{B_s}$, which are then controlled by  $(g_{Q_{D}^3}/t)^2$ according to Eq.~\eqref{gijAL}. Therefore contraints from $B-\bar B$ mixing restrict the degree of $Q_U^3$ compositeness, thus introducing some tension between $B$ physics and top $A_{\rm FB}$. This tension is illustrated in Fig.~\ref{fig_flavor-afb} where $\delta A_{\rm FB}^{450}$ is plotted as a function of $c_{Q^3}$ (assuming benchmark values for the other relevant parameters). 
For instance, the region with $c_{Q^3}<0$ leading to $\delta A_{\rm FB}^{450}\gtrsim10\%$, is excluded by flavor physics.
%

\section{Dijet searches at the LHC}\label{sec:dijets}
\begin{figure}[tb]
\begin{center}
\includegraphics[width=0.8\textwidth]{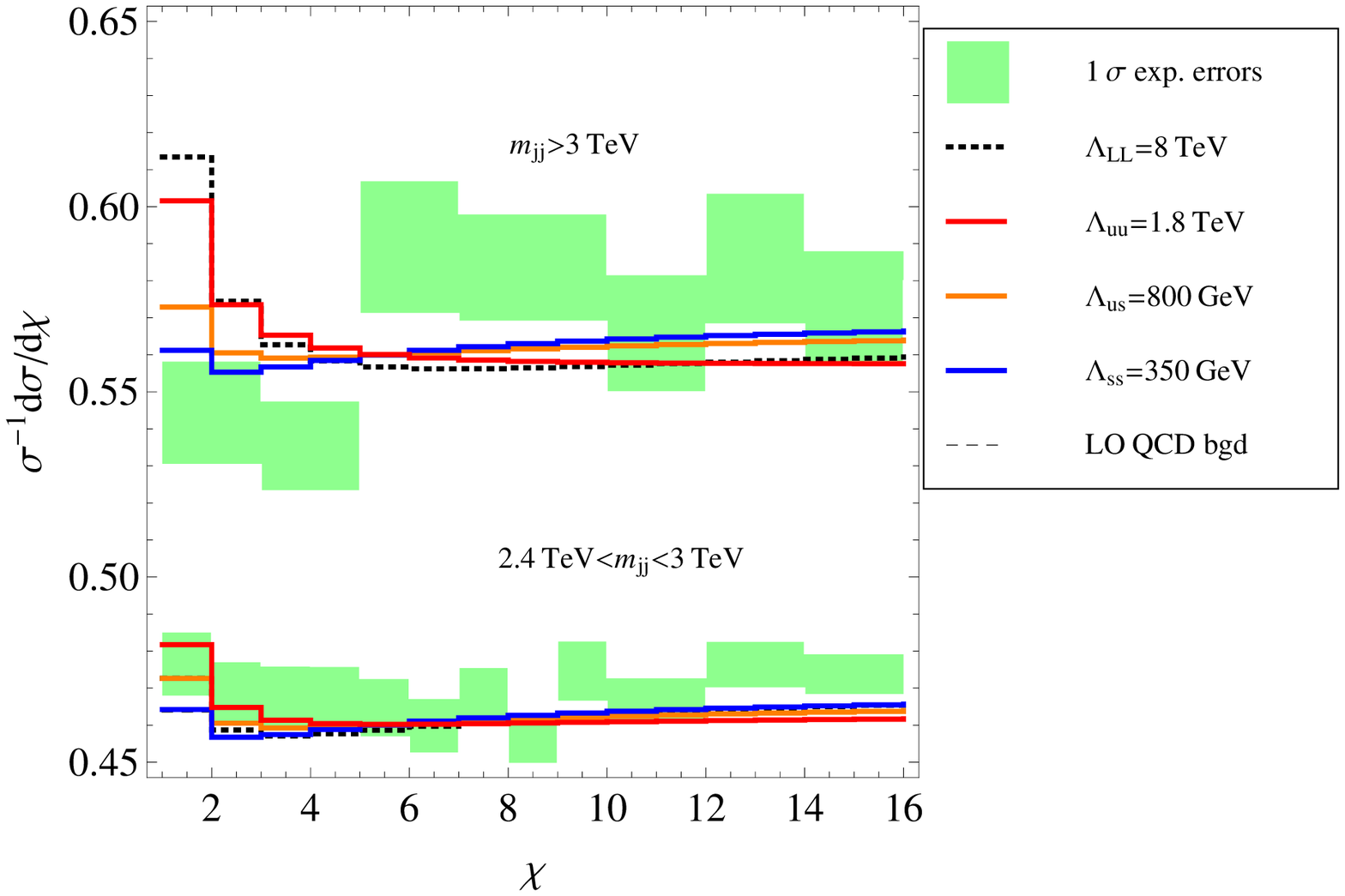}
\end{center}
\caption{$d\sigma/d\chi$ distributions as analysed by CMS~\cite{CMSdijets} at the 7$\,$ TeV LHC for $2.4\,{\rm TeV}<m_{jj}<3\,{\rm TeV}$ (down) and $m_{jj}>3\,$TeV (top); A shift  on the $y$-axis of $+0.4$ and $+0.5$, respectively, is understood. The red, blue and orange lines include the contribution of the contact interactions in Eq.~\eqref{EFTjj} at a scale of $(\Lambda_{uu},\Lambda_{us},\Lambda_{ss})=(1.8,0.8,0.35)\,$TeV respectively with negative Wilson coefficients, while the dashed black line is the QCD only LO expectation. For the sake of comparison the dotted black lines are the predictions from the operator $+2\pi/\Lambda_{LL}^2(\bar q_L \gamma^\mu q_L)^2$ with $q_L$ the first generation LH doublet whose scale is set to the $95\%$CL bound reported by CMS $\Lambda_{LL}=8\,$TeV. The green shaded areas denote the experimental errors at one standard deviation.}
\label{figdijetCMS}
\end{figure}

Dijet production is a simple channel to look for NP at hadron colliders. In particular, dijet searches are sensitive to NP states which couple sizably to light quark flavors (or gluons). Therefore these searches are typically powerful in constraining models devised for the top $A_{\rm FB}$ whose large measured value does require non-negligible NP couplings to up (or down) quarks. In cases where the NP states are heavy enough to avoid on-shell production at colliders, dijet searches can still be  constraining indirectly this NP by studying the dijet angular distribution. The reason is that heavy NP (not necessarily off-shell) produces relatively isotropic dijets as compared to QCD which preferentially produces dijets in the forward regions. A useful angular variable employed to discriminate NP dijets from the QCD background is $\chi\equiv e^{2y}$, where $y$ is the jet rapidity in the partonic center-of-mass frame. This variable is useful in that the QCD background is approximately evenly distributed in $\chi$, while the heavy NP events populate the low $\chi$ region as shown on Fig.~\ref{figdijetCMS}. Since the present dijet angular distribution is consistent with the SM shape the experimental data put bounds on the heavy NP interactions involving light quarks. The CMS analysis is based directly on the $d\sigma_{jj}/d\chi$ distributions for various bins of dijet invariant masses~\cite{CMSdijets}. ATLAS uses a slightly different angular function $F_\chi(m_{jj})\equiv \sigma_{jj}(\chi<3.3)/\sigma_{jj}(\chi<30)$, which exploits the same kinematical difference but by measuring the ratio of central to forward dijet events for various invariant masses~\cite{ATLASdijets}. We evaluate below the dijet angular distribution for the warped model defined in Section~\ref{section_model}, as well as for the set of effective operators arising in models where the NP states responsible for the top and charm anomalies is not directly produced at the $7\,$TeV LHC. All cross-sections are computed at the partonic level and at leading order in the QCD coupling.

\begin{figure}[tb]
\begin{center}
\includegraphics[width=0.6\textwidth]{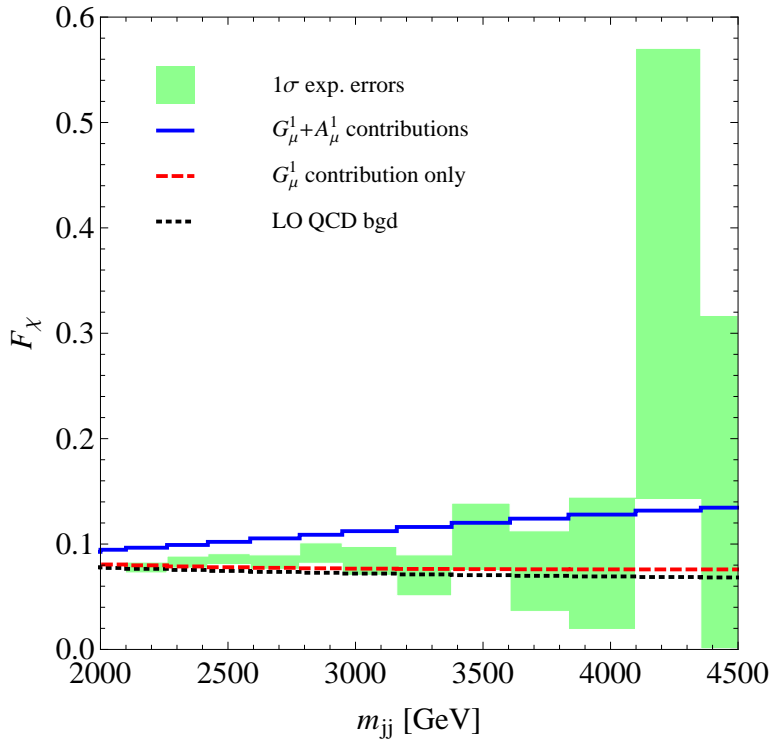}
\end{center}
\caption{$F_\chi$ variable as a function of the dijet invariant mass $m_{jj}$ as analysed by ATLAS~\cite{ATLASdijets} at the 7$\,$ TeV LHC. The solid blue (dashed red) line is the distribution predicted by the model defined in Section~\ref{section_model} (without) including the first axial resonance, while the dotted black like corresponds to the QCD background. The green shaded areas denote the experimental errors at one standard deviation.} 
\label{figdijetATLAS}
\end{figure}

We argued in Sections~\ref{AFB} and~\ref{section_CPV-LHCb} that accomodating the  top and charm anomalies through four-fermi contact interactions in partially composite models requires some of the light quark flavors to be rather composite. In particular the $s_R$, $c_R$ and $u_R$ quarks have large couplings to color octet KK resonances, potentially in tension with dijet searches. The RH up quark compositeness is most severely constrained through $uu\to uu$ transitions. Dijet searches are less sensitive to the RH strange and charm quark compositeness since these quarks have suppressed PDFs relative to the (valence) up quark. We show in Fig.~\ref{figdijetATLAS} the angular distribution of dijet at the LHC running at $7\,$TeV as predicted by the warped model defined in Section~\ref{section_model}. 
The model appears in tension with the current dijet data, especially in the $2\,$TeV$\lesssim m_{jj}\lesssim 3\,$TeV region. 
We stress however that the ignored next-to-leading order corrections tend to reduce the NP contributions to the dijet cross-section by $\sim \mathcal{O}(20\%)$~\cite{CPYuan,CMSdijets}. Moreover, a large fraction of NP events are $s\bar s$ dijets (mediated by the axial KK gluon) which might enjoy lower detector's acceptance. Both effects might relax the tension with the current dijet data.

We now present a more generic analysis which is relevant to NP models where the new states responsible for the top and charm anomalies are heavy enough not to be produced on-shell at the $7\,$TeV LHC. In this case, the dominant NP contributions to dijet production  arises from following effective Lagrangian
\beq\label{EFTjj}
\mathcal{L}_{\rm eff}^{jj}= \frac{c_{uu}^8}{\Lambda_{u u}^2}\mathcal{O}_{uu}^8+\frac{c_{ ss}^8}{\Lambda_{ ss}^2}\mathcal{O}_{ ss}^8+\frac{c_{us}^8}{\Lambda_{us}^2}\mathcal{O}_{us}^8\,,
\eeq
where 
\bea
\mathcal{O}_{uu}^8&=&\frac{1}{2}(\bar{u}_R \gamma^\mu T^a u_R)^2\,,\\
\mathcal{O}_{ss}^8&=&\frac{1}{2}(\bar{s}_R \gamma^\mu T^a s_R)^2\,,\\
\mathcal{O}_{us}^8&=&(\bar{u}_R \gamma^\mu T^a u_R)(\bar{s}_R \gamma_\mu T^a s_R)\,.
\eea
We normalize the Wilson coefficients in Eq.~\eqref{EFTjj} as $c_i=\pm 1$. Contact interactions involving the charm quark field would also arise. The suppressed charm quark PDF relative to up quarks together with the relatively low degree of compositeness required by the charm CPV anomaly makes its contributions to dijet production subdominant. We thus ignore four-fermion interactions involving the charm quark.
We assume that NP state giving rise the Lagrangian Eq.~\eqref{EFTjj} is exchanged in the $s$-channel so the relation between the operator scales $\Lambda_{us}=\sqrt{\Lambda_{uu}\Lambda_{ss}}$  holds and $c_{us}^8=c_{uu}^8=c_{ss}^8=-1$. We calculated analytically, up to $\mathcal{O}(1/\Lambda^4)$, the relevant partonic matrix elements which arise from the Lagrangian Eq.~\eqref{EFTjj}\footnote{Our matrix element expressions agree with those of Ref.~\cite{Jarlskog:1990bk}. Notice that some of the matrix elements in Refs.~\cite{SuperCol} are erroneous since inconsistent with the crossing symmetry.} and we used MSTW~\cite{MSTW} parton distribution functions (PDFs) at LO to compute the hadronic cross-sections leading to dijet production.
As argued in Section~\ref{AFB} the measured $A_{\rm FB}$ requires a large coupling of RH up quarks to the NP resonance. We thus assume that the $u_R$ compositeness scale saturates the bound from current dijet searches. Since neither CMS nor ATLAS provided a full analysis of the operator $\mathcal{O}_{uu}^8$ we derive an approximate bound on the latter by requiring that the shape of the angular distribution is not more distorted than in the presence of the operator $+2\pi/\Lambda_{LL}^2(\bar q_L \gamma^\mu q_L)^2$, with $q_L$ the first generation LH quark doublet, whose scale is set to the $95\%$CL bound reported by CMS $\Lambda_{LL}=8\,$TeV. We find $\Lambda_{uu}\gtrsim 1.8\,$TeV for $c_{uu}^8=-1$. This bound is in agreement with the result of Ref.~\cite{Pomarol}. As argued above the charm seesaw between the $\Delta a_{CP}$ measurement and the bound on $D-\bar D$ mixing requires $\Lambda_{ss}\lesssim 350\,$GeV. Dijet invariant masses up to $m_{jj}\simeq 3\,$TeV were probed so far at the LHC. We stress that the integrated out NP must have a strong, yet still pertubative, coupling to RH strange quarks in order to induce the $\mathcal{O}_{ss}^8$ operator at such a low scale. We plot in Fig.~\ref{figdijetCMS} the angular distribution analysis by CMS which results from the contact interactions Eq.~\eqref{EFTjj} set at the scales $\Lambda_{uu}\simeq 1.8\,$TeV, $\Lambda_{ss}\simeq 350\,$GeV and hence $\Lambda_{us}\simeq 800\,$GeV. Although the scales of $\mathcal{O}_{us}^8$ and $\mathcal{O}_{ss}^8$ are low their contributions to dijet production are still suppressed enough thanks to the suppressed strange quark PDF. We conclude from this analysis that for heavy NP not accessible at the $7\,$TeV LHC the parameter space accomodating top $A_{\rm FB}$ and $\Delta a_{CP}$ is consistent with the present dijet data.\\

\section{Other collider signatures}\label{othersignal}

\subsection{$t\bar{t}$ spectrum at the LHC}
\begin{figure}[tb]
\begin{center}
\includegraphics[width=.6\textwidth]{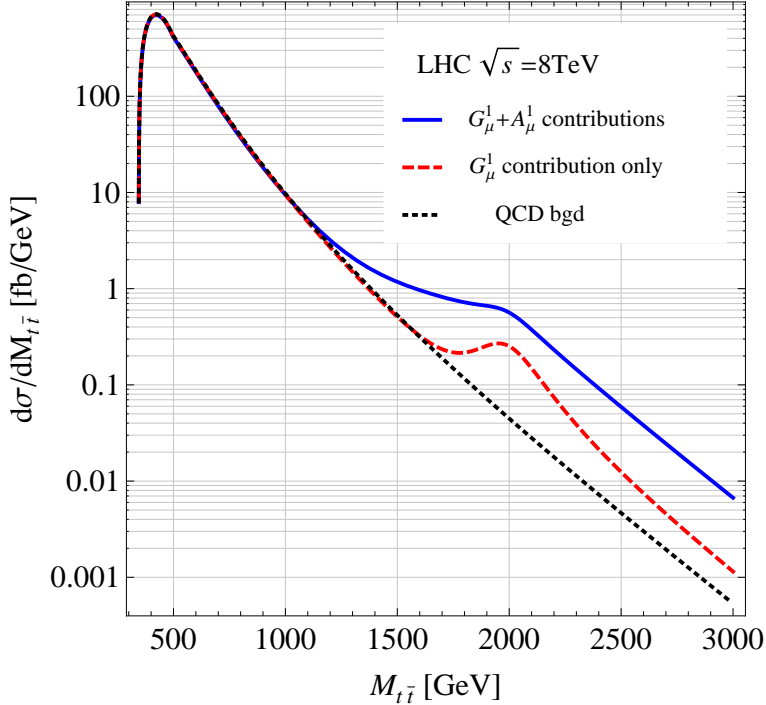}
\end{center}
\caption{Differential cross section for top pair production at the LHC 8 TeV as predicted by the benchmark parameters in  Eqs.~\eqref{bulkC}-\eqref{benchmark}. The solid blue line is the SM+NP $m_{t \bar t}$ distribution  when both the vector and axial KK gluons are present, while the dashed red line shows the spectrum obtained by the SM and the vector KK gluon only ($m_A\gg m_V$); the dotted black line is the SM QCD expectation alone.} 
\label{ttbar-lhc}
\end{figure}

We discuss here the signature of the vector and axial KK gluon resonances for the differential cross-section of $t\bar t$ production at the LHC. Figure~\ref{ttbar-lhc} shows the expected spectrum as predicted by the benchmark parameters in  Eqs.~\eqref{bulkC}-\eqref{benchmark} for the current LHC run at 8 TeV. The vector gluon has a relatively narrow width of $\sim 0.2m_V$, whereas the axial KK gluon is typically much broader with a width of $\sim 0.7m_A$ which arises mostly from  its large couplings to the  first composite light RH quarks as well as to $t_L$ and $b_L$. Since the two resonances are quite degenerate, such a broad axial gluon masks completely the bump the narrower vector gluon would leave in the $t\bar t$ spectrum. Such a broadly spread NP signal will probably be hard to pick with resonance searches techniques developed for boosted tops~\cite{MarcelVos}. Even if so our $t\bar t$ signal may still probably be tested by measurements of the integrated cross-section tail above a certain high invariant mass. For instance, we find that integrating our predicted distribution above $m_{t\bar t}>1\,$TeV yields an ehancement factor of $\simeq 1.5$ over the SM cross-section. Such a measurement was actually already carried by the CMS experiement in their all hadronic boosted top sample~\cite{allhadCMS} where they bounded such an ehnancement to be $< 2.6$ at $95\%$CL. Our signal may be in reach of LHC 8 TeV.

\subsection{Bottom and lepton-based top $A_{\rm FB}$ at the Tevatron}
\begin{figure}[tb]
\begin{center}
\includegraphics[width=.8\textwidth]{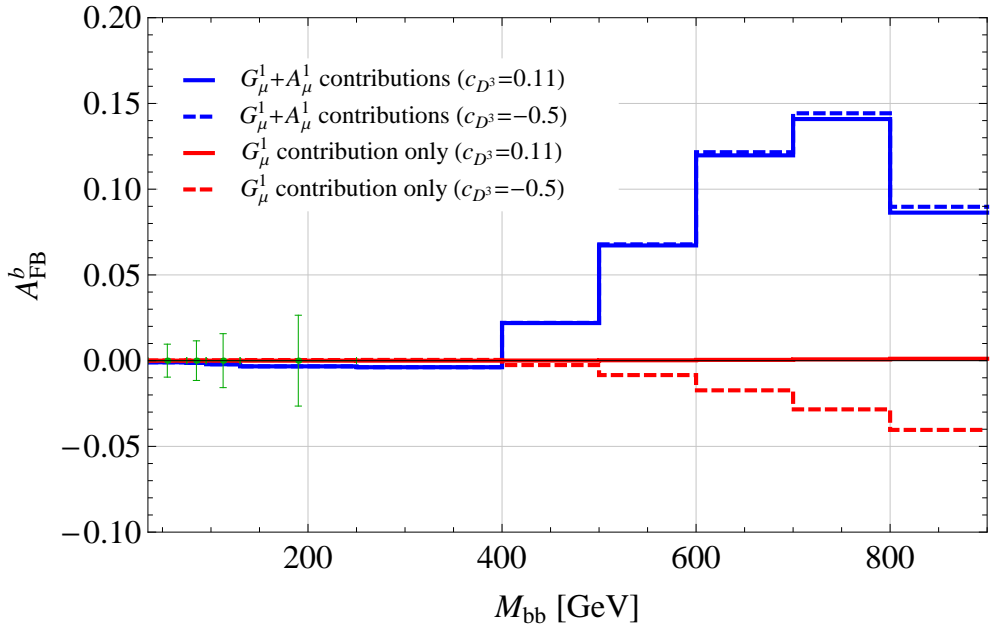}
\end{center}
\caption{$A_{\rm FB}$ in $b\bar{b}$ production at the Tevatron as predicted by the model defined in Section~\ref{section_model}. Blue lines show the asymetry arising in the case where both the vector and axial KK gluons are present in the spectrum, while red ones are the predictions of the vector KK gluon only ($m_A\gg m_V$). Dashed curves show corresponding predictions assuming a fully composite $b_R$ ($c_{D^3}\simeq -0.5$). The green bars illustrate the expected sensitivity of the forthcoming CDF measurement at low $m_{b\bar b}$ (see {\it e.g.} Ref.~\cite{CDFbb}).} 
\label{bottom-afb}
\end{figure}

The dominant production mechanism for top $A_{\rm FB}$ is through $\bar u_R u_R \to \bar{t}_L t_L$. This is a direct consequence of the EWPTs which forces the RH tops to be rather elementary. Since LH tops are dominantly produced a potentially sizable charge asymmetry in bottom pair production is also expected. Such a measurement is currently performed by the CDF collaboration which the expected sensitivity was reported in Ref.~\cite{CDFbb}. We show on Fig.~\ref{bottom-afb} the prediction of our model for the benchmark parameters of Eqs.~\eqref{bulkC}-\eqref{benchmark}. For $m_{b\bar b}\lesssim 200\,$GeV, which is the regime to be probed first by CDF the expected $A_{\rm FB}$ is rather small, typically less than $\mathcal{O}(1\%)$. This small value is mostly due the fact that the $\mathcal{O}(2\,$TeV$)$ vector and axial KK gluons are almost totally decoupled at these low energies. An interestingly large asymmetry of $\mathcal{O}(10\%)$ is however predicted at invariant masses above $m_{b\bar b}\gtrsim 600\,$GeV, a regime possibly in reach by the Tevatron experiments. Furthermore we  show that most of this NP asymmetry in bottom pair production is supported by the axial state.
\\

The forward-backward charged lepton (or lepton-based) asymmetry  in $t\bar t$ production at the Tevatron is a useful discriminant in order to better characterize the NP source explaining to the top $A_{\rm FB}$~\cite{AFBlep,FPS,AFBlep2}. In particular, the lepton-based asymmetry near the $t\bar t$ threshold directly probes the chiralities of the initial $q\bar q$ pair~\cite{FPS}. Since NP $A_{\rm FB}$ contributions here originate from RH up quark/antiquark collisions, a negative contribution to the lepton-based asymmetry is expected near threshold. At higher $t\bar t$ invariant masses ($m_{t\bar t}\gtrsim 1\,$TeV) the produced tops become ultra-relativistic and valuable informations about their chiralities (or helicities) can be inferred from the kinematics of their decay products. In particular, when approaching the KK gluon resonances, we expect from the produced LH top pairs a softer lepton and a harder $b$-jet $p_T$ spectra relative to the SM $t\bar t$ production~\cite{TOPpol}.

\section{Conclusions}\label{conc}

We showed that the top forward-backward and charm CP asymmetries can be accommodated conjointly within warped extra dimension or partial compositeness models, provided RH quarks of the first two generations are composite fields. This feature constitutes a common microscopic origin for both asymmetries in this framework. It is remarkable that the two observables can be explained through sizable NP contributions without conflicting with the colossal amount of data currently at disposal on flavor physics, EWPTs, dijet and $t\bar t$ resonance searches at the LHC. Yet, the flavor ``paradigm'' advocated in this paper is very predictive. If indeed both top $A_{\rm FB}$ and charm $\Delta a_{CP}$ originate from NP, we showed that the following predictions are soon to be tested by the LHC experiments.
First of all, since the observed charm direct CPV is induced through a four-fermion operator of the form $(\bar u c)_{V+A}(\bar s s)_{V+A}$ a significantly larger CP asymmetry in the $D\to K^+K^-$ rate than in the $D\to \pi^+\pi^-$ one is expected. This is in contrast with both SM-like and chromomagnetic dipole-based NP explanations which predicts $a_{CP}^{\pi\pi}\gtrsim a_{CP}^{KK}$. Measurements of the individual rate asymmetries (and also possibly of other radiative decays~\cite{raddecays}) at flavor factories will provide a clear test of the type of contributions leading to the now well established CPV effect reported in $\Delta a_{CP}$.
The $\mathcal{O}(10\,)$TeV scale of the above four-fermion operator requires both a saturated NP bound on CP conserving $D-\bar D$ mixing and composite (RH) strange quark. $\Delta a_{CP}$ implies sizable CPV in $D-\bar D$ mixing (assuming a maximal CPV phase in the up sector) as well as an excess of dijets at the LHC. Both effects are expected to be observed by forthcoming measurements at the LHC.
The top $A_{\rm FB}$ is induced through tree-level exchange of KK gluon resonances. If the SM expectation is not significantly enhanced by NNLO QCD corrections, we argued that the large asymmetry reported by Tevatron's experiments cannot be accomodated in composite Higgs models, unless {\it e.g.} the color symmetry is extended to SU(3)$_L\times$SU(3)$_R$ in the strong sector, thus providing an axial resonance which can significantly contribute to the top $A_{\rm FB}$. We also showed that the axial resonance, whose couplings to quarks reproduce the observed top asymmetry, induces a negative contribution to the lepton-based $A_{\rm FB}$ in $t\bar t$ events and an $\mathcal{O}(10\%)$ $A_{\rm FB}$ for $m_{b\bar b}\gtrsim 600\,$GeV in bottom pair production at the Tevatron, as well as an excess of $t\bar t$ pairs relative to the SM potentially visible at the LHC $8\,$TeV run.

\section*{Acknowledgments}
We thank Kaustubh Agashe, Gino Isidori and Yotam Soreq for useful discussions. LD thanks the Theory Division of CERN Physics Department for its hospitality while part of this work was completed.
LD is partly supported by FONCYT under the contract PICT-2010-1737,
CONICET under the contract PIP 114220100100319 and EPLANET
under the contract PIRSES-2009-GA-246806. The work of CG is partly supported by the European Commission under the contract ERC advanced
grant 226371 MassTeV and the contract PITN-GA-2009-237920 UNILHC.
GP is the Shlomo and Michla Tomarin development chair, supported by the grants from GIF, Gruber foundation, IRG, ISF and Minerva.

\appendix

\section{Boundary conditions of 5D fermion fields}\label{BCpsi}

We describe the boundary conditions (BCs) on the 5D fermion fields defined in Eqs. (\ref{embedding12}) and (\ref{embedding3}) which lead to the SM chiral spectrum in 4D. 
We focus only on the LH sectors where the EW representations of the 5D fermions are larger than the corresponding SM representations in 4D. The extra chiral zero modes are removed from the low energy spectrum by mean of the following BCs for all three generations~\footnote{We only write the BCs for the Weyl spinors sharing the same chirality as the zero mode. The KK partners of opposite chirality have opposite BCs, see {\it e.g.} Ref.~\cite{NeuGross}.}
\beq\label{BCs}
Q_{U}=\left(\begin{array}{cc}
u_L^1\ [++] & X_L\ [-+] \\
d_L^1\ [++] & U_L\ [-+] 
\end{array}\right),\  
Q_D=\left(\begin{array}{cc}
D_L\ [-+] & u_L^2\ [++] \\
S_L\ [-+] & d_L^2\ [++] 
\end{array}\right),\  
\eeq
\beq
L=\left(\begin{array}{cc}
\nu_L\ [++] & F_L\ [-+] \\
e_L\ [++] & N_L\ [-+] 
\end{array}\right)\,,
\eeq
where a $+$ ($-$) denotes a Neumann (Dirichlet) BC and the first (second) sign corresponds to the UV (IR) brane. Only the lower case components, which satisfy a Neumann BC on both branes, have a zero mode identified to the chiral SM fermions.
Notice that $Q_{U,D}$ and $L$ satisfy Neumann BC on the
IR brane, thus preserving the full EW bulk symmetry. On the UV brane, BCs are Neumann for components with hypercharge
$Y\equiv T^{3}_R+X=1/6$ for $Q_{U,D}$ and $Y=-1/2$ for $L$, while all other components have Dirichlet BCs. This explicitly breaks
SU(2)$_R\times$U(1)$_X$ but preserves SU(2)$_L\times$U(1)$_Y$ in the UV. 

The BCs in Eq.~\eqref{BCs} lead however to two LH quark zero mode doublets
per generation, $q_L^1\equiv(u_L^1,d_L^1)$ and $q_L^2\equiv(u_L^2,d_L^2)$. In order to decouple the extra quark doublets we introduce, for each generation, one
4D RH doublet $\chi_R$ localized on the UV brane, which marries a
linear combination of $q_L^{1,2}$ through a heavy mass $m_{\rm UV}\sim
k$ as
\beq
\mathcal{L}_{\rm UV}\supset m_{\rm UV} \bar \chi_R
(q_L^1\sin\alpha-q^2_L\cos\alpha)|_{z=R} + \rm{h.c.}\,,
\eeq
For simplicity we assume a trivial flavor structure for $m_{\rm
UV}$ and the angle $\alpha$. The linear combination $(q_L^1\sin\alpha -q_L^2\cos\alpha)$ of would-be zero mode doublets
decouples at low energies, while the orthogonal combination remains massless and is identified with the SM quark doublet.

\section{Vector and axial KK gluon properties}

\subsection{KK decompositions and couplings to SM-like quark zero modes}\label{KKdefs}

We present here the KK decompositions of the gauge and fermion 5D fields which allow to derive the couplings relevant for the phenomenology discussed in the main text. We work with the 5D metric  defined in Section~\ref{section_model}. All 5D
fields are decomposed as
\begin{equation}
\Phi(x,z)=\sum_n \psi_{\Phi^n}(z) \Phi^{n}(x)\,,
\end{equation}
where $\Phi^{n}(x)$ are 4D KK modes and $\psi_{\Phi^n}(z)$ are the corresponding wavefunctions characterizing the KK mode's profiles
along the extra dimension. The wavefunctions are obtained by solving the bulk equations of motion (EOM) for the 5D field $\Phi(x,z)$ (see {\it e.g.} Ref.~\cite{GhergPom}). 

If a 5D fermion field $f$ has Neumann boundary conditions both on the
UV and IR branes, there is a 4D chiral zero mode whose wavefunction is 
\begin{equation}
\psi_{f^0}(c_f,z)= R^{\prime3/2}k^2\left(\frac{z}{R'}\right)^{2-c_f} \chi_f \,,\quad \chi_f=\sqrt{\frac{1-2c_f}{1-\exp[-\xi(1-2c_f)]}}\,,
\end{equation}
which satisfies the following normalization condition\footnote{We neglect  potential kinetic terms localized on the branes throughout.}
$\int_R^{R'}
dz\psi_{f^0}(c_f,z)^2/(kz)^4=1$. $c_f$ is the fermion bulk mass (expressed in units of $k$) and $\xi \equiv \log(R'/R)$. Similarly, the 5D scalar field $S$ and
gauge field $V_\mu$ zero modes are\footnote{The fifth component of 5D gauge field $V_5$ yields a set of 4D scalar modes. For boundary conditions allowing for a massless 4D gauge zero mode, a massless zero mode for $V_5$ is forbidden. Moreover all KK modes of $V_5$ are not physical and can be identified to the longitudinal polarizations of the 4D gauge KK modes~\cite{GaugeInterval}.}
\begin{equation}\label{SVzeromodes}
\psi_{S^0}(\beta_S,z)= \left(\frac{2(1+\beta_S)}{1-e^{\xi(2+2\beta_S)}}\right)^{1/2}\left(\frac{z}{R'}\right)^{2+\beta_S}
R'k^{3/2} \,, \quad \psi_{V^0}(z)=\left(\frac{k}{\xi}\right)^{1/2} \,,
\end{equation}
respectively. $\beta_S\equiv\sqrt{4+m_S^2/k^2}$, where is $m_S$ the scalar field's bulk mass. The wavefunctions are normalized as $\int_R^{R'}
dz\psi_{S^0}(z)^2/(kz)^3=1$ and $\int_R^{R'} dz\psi_{V^0}(z)^2/(kz)=1$. 

Wave
functions of the gauge field KK modes are
\begin{equation}\label{Awf}
\psi_{V^n}(z)=\frac{z}{N^V_n}[J_1(m^V_nz)+b^V_nY_1(m^V_nz)]  \,,
\end{equation}
where $J_p(x)$ and $Y_p(x)$ are order $p$ Bessel functions of the first and second kind, respectively. $N^V_n$ are constants set to 
\beq\label{NAwf}
(N_n^V)^2=\frac{1}{2}\left[R^{\prime2}\left(J_1(m_n^VR')+b_n^VY_1(m_n^VR')\right)^2-R^{2}\left(J_1(m_n^VR)+b_n^VY_1(m_n^VR)\right)^2\right]
\eeq
by the orthonormality condition $\int_R^{R'} dz \psi_{V^n}\psi_{V^m}/(kz)=\delta_{nm}$.
The constants 
$b^V_n$ and the KK masses $m^V_n$ depend on the boundary conditions. For 5D field satisfying Neumann conditions on both boundaries (such as SU(3)$_V$ gauge fields) they are set by
\beq\label{A++}
\frac{J_0(m_n^VR)}{Y_0(m_n^VR)}=\frac{J_0(m_n^VR')}{Y_0(m_n^VR')}=-b_n^V\,.
\eeq
In particular the lightest KK mass is $m^V_1\simeq 2.45/R'\equiv m_V$.
On the other hand for 5D fields satisfying a Dirichlet condition on the UV brane and a Neumann one on the IR brane (such as the axial gauge fields of the coset SU(3)$_L\times$SU(3)$_R/$SU(3)$_V$), one has
\beq\label{A-+}
\frac{J_1(m_n^VR)}{Y_1(m_n^VR)}=\frac{J_0(m_n^VR')}{Y_0(m_n^VR')}=-b_n^V\,.
\eeq
Notice that the above KK wavefunctions are obtained in the absence of spontaneous symmetry breaking (SSB) effects. 
In the case of the axial KK modes living in SU(3)$_L\times$SU(3)$_R/$SU(3)$_V$, an additional source of breaking arises from the $\phi\sim({\bf 3, \bar 3})$ scalar VEV which modifies the quadratic part of the 5D gauge fields EOM and thus affects the wavefunctions of the axial KK modes. There is in general no analytic solution for the KK wavefunctions in this case and one has to solve the bulk EOM numerically~\cite{DaRold:2005zs}. However, in the limit where the SSB scalar VEV $v_\phi$ is much smaller than the lightest KK mass, the former can be treated perturbatively and the above expressions for the KK gauge wavefunctions are a good approximation to the exact ones up to corrections of $\mathcal{O}(v_\phi^2/m_V^2)$.\\

The 5D gauge fields $G_\mu$ of SU(3)$_V$ and $A_\mu$ of SU(3)$_L\times$SU(3)$_R/$SU(3)$_V$ are related to the original 5D fields $A^L_\mu$ and $A^R_\mu$ associated with the bulk symmetry SU(3)$_L$ and SU(3)$_R$, respectively, as 
\beq
G_\mu= c A^L_\mu + s A^R_\mu\,, \quad A_\mu = -sA^L_\mu + c A_\mu^R\,, 
\eeq
where $c\equiv g_R/\sqrt{g_L^2+g_R^2}$ and $s\equiv g_L/\sqrt{g_L^2+g_R^2}$, with $g_L$ ($g_R$) the 5D gauge coupling in units of $k^{-1/2}$ associated with the SU(3)$_L$ (SU(3)$_R$) bulk symmetry. The 5D fields $G_\mu$ and $A_\mu$ are associated with the 5D gauge couplings $g_Lg_R/\sqrt{g_L^2+g_R^2}$ and $\sqrt{g_L^2+g_R^2}$, respectively, and they have the following KK decompositions
\beq
G_\mu(x,z)= \psi_{V^0}(z) g_\mu(x) + \sum_{n\geq1} \psi_{G^n}(z)G_\mu^{n}(x)\,,\quad
A_\mu(x,z)=\sum_{n\geq1}\psi_{A^n}(z)A_\mu^{n}(x)\,,
\eeq
where the zero mode $g_\mu(x)$ is identified with the QCD massless gluon, whose wavefunction is $\psi_{V^0}(z)$ is constant and given at tree-level by Eq.~\eqref{SVzeromodes}. The KK wavefunctions $\psi_{G^n}(z)$ ($\psi_{A^n}(z)$) are given by Eqs.~\eqref{Awf}-\eqref{NAwf} and Eq.~\eqref{A++} (Eq.~\eqref{A-+}).

One combination of the two 5D gauge couplings of the color sector is set by the 4D gauge coupling of QCD $g_s(\mu\simeq 1/R')\simeq 1$ through the following matching condition~\cite{gmatching}
\begin{equation}
\frac{1}{g_s^2}=\xi\left(\frac{1}{g_V^2}+\frac{b_{QCD}}{8\pi^2}\right) +\frac{1}{g_{UV}^2}+\frac{1}{g_{IR}^2}\,,
\end{equation}
where 
\beq
g_V\equiv \frac{g_L g_R}{\sqrt{g_L^2+g_R^2}}\,,
\eeq
and $b_{QCD}=-7$ is the one-loop coefficient of the $\beta$-function to which only elementary fields contribute to and $g_{UV}$ ($g_{IR}$) denotes eventual gauge kinetic term contributions on the UV (IR) brane. In the absence of brane gauge kinetic terms ($g_{UV}=g_{IR}\to \infty$), one finds $g_V\simeq 3$ at one-loop~\cite{Agashe2site}. The higher value of $g_V$ in Eq.~\eqref{benchmark} can be obtained through {\it e.g.} adding a UV kinetic term of $g_{UV}\simeq 0.7$ (leaving $g_{IR}\to \infty$).
The other combination of 5D color gauge couplings $t\equiv g_L/g_R$ is left as a free parameter.

Couplings between gauge KK modes and the quark zero modes are obtained by performing 
 overlap integrals over the extra dimension, which involve the appropriate wavefunctions and metric factors. 
The coupling of first SU(3)$_V$ KK mode $G^1_\mu$ to the chiral zero-mode fermion current $\bar f\gamma^\mu f$ is 
\beq
g_{f}^V\equiv g_V\int_R^{R'} dz\,(kz)^{-4}\psi_{f^0}(c_f,z)^2\psi_{V^1}(z)\,,
\eeq
while the first axial KK mode $A_\mu^1$ coupling to zero-mode fermions is 
\beq
g_{f}^A\equiv g_V s(t) \int_R^{R'} dz\,(kz)^{-4}\psi_{f^0}(c_f,z)^2\psi_{A^1}(z)\,.
\eeq 
The above KK couplings are well approximated by~\cite{CsakiFalkowskiWeiler} 
\beq 
g_{f}^V\simeq g_V\left[\chi_f^2\gamma_f-\xi^{-1}\right]\,,\quad g_f^A\simeq g_f^V s(t)\,,
\eeq
where $\gamma_f \equiv \frac{\sqrt{2}x_1}{J_1(x_1)}\int_0^1 dx x^{1-2c_f}J_1(x_1 x)\simeq 2.3/(3-2c_f)$, $x_1\simeq 2.4$ being the first zero of the $J_0$ Bessel function, $J_0(x_1)=0$. 
\subsection{Vector/axial mass splitting}\label{splitting}

We observe that $\delta m^2\equiv m_A^2-m_V^2$ receives sizable contributions from the necessary breaking of SU(3)$_L\times$SU(3)$_R$ in the IR. This contribution to the mass splitting is parametrically larger than the one arising from the breaking on the UV boundary. We estimate the mass splitting when the IR breaking is realized by the VEV of an EW singlet $\phi\sim({\bf 3,\bar 3})$, as assumed in the text.
We parameterize the $\phi$ VEV as $\langle\phi\rangle=v_\phi \lambda^0$, with $\lambda^0\equiv{\bf 1}_{3\times3}$, and we pursue a perturbative treatment of the latter assuming $v_\phi\lesssim m_V$. At leading order in VEV insertion, we find
\beq
\delta m^2\simeq (g_L^2+g_R^2)v^2_\phi\,\mathcal{O}_A(\beta_\phi). 
\eeq
$\mathcal{O}_A(\beta_\phi)$ is the overlap integral between the colored scalar VEV and the axial KK-gluon, which is defined as 
\begin{equation}
\mathcal{O}_A(\beta_\phi)=\int_{R}^{R'} dz (kz)^{-3}\,\psi_\phi(\beta_\phi,z)^2 \psi_{A^1}(z)^2\,,
\end{equation}
where $\psi_\phi(\beta_\phi,z)$ and $\psi_{A^1}(z)$ are the colored scalar VEV and axial KK-gluon profiles, respectvely, and
$\beta_\phi$ controls the localization of the VEV profile along the extra dimension. We find  $\mathcal{O}_A(\beta_\phi)=2$ for an exactly IR-brane localized scalar, while $\mathcal{O}_A(\beta_\phi)\simeq 1$ for $\beta_\phi=0$. We argue that the top mass yields a lower bound on the $\phi$ VEV's size. Following notations of Section~\ref{section_model}, the top mass is approximately $m_t\sim Y_* v(v_\phi/\Lambda) \chi_{Q^3}\chi_{U^3}/\sqrt{2}$ where $v\simeq 246\,$GeV is the SM-like Higgs VEV and $Y_*$ is the anarchic Yukawa. $\Lambda\sim N_{\rm KK}k$ is the 5D cut-off scale, where $N_{\rm KK}$ is the number of perturbative KK states in the low-energy effective theory. Naive dimensional analysis~\cite{NDA} (NDA) estimates the maximal perturbative Yukawa to be $Y_*\lesssim 16\pi^2/\sqrt{N_c}$ which, in turn, implies 
\beq
\frac{v_\phi}{\Lambda}\gtrsim 10^{-2}\left(\frac{16\pi^2/\sqrt{N_c}}{Y_*}\right)
\eeq
assuming a fully composite top, {\it i.e.} $\chi_{Q^3}\simeq \chi_{U^3}\sim\mathcal{O}(1)$. The gauge couplings $g_{L,R}$ are related to the 5D vector KK-gluon coupling $g_V$ as $g_L^2+g_R^2 = g_V^2(1+t^2)/t$. We thus find the following estimate for the axial/vector KK mass splitting
\beq
\frac{\delta m^2}{m_V^2}\simeq g_V^2(t+t^{-1})N_{\rm KK}^2\left(\frac{v_\phi}{\Lambda}\right)^2 \gtrsim 0.1\,,
\eeq
assuming $N_{\rm KK}=3$, $t\simeq 1$ and a maximal gauge coupling of $g_V\sim 4\pi/\sqrt{N_{\rm KK}}\simeq 7$. The parameters' choice in Eq.~\eqref{benchmark} is consistent with the above estimate.

\section{Kaon physics in anarchic SU(3)$_L\times$SU(3)$_R$ models.}\label{App_kaon}

We consider the flavor phenomenology of warped models based on the extended bulk gauge symmetry SU(3)$_L\times$SU(3)$_R$. In contrast with the peculiar flavor ansatz assumed in the text, we address below the status of this type of models in the limit where the fundamental flavor parameters are anarchic. We assume that all LH and RH SM fermions are embedded in $({\bf 3,1})$ and $({\bf 1,3})$ of the extended color group, respectively. We focus on both direct and indirect CPV observables in Kaon system as they provide the most severe constraints to anarchic warped/composite models where the color bulk gauge symmetry is simply SU(3)$_V$. We refer below to this type of models as arnarchic RS. The above extended color symmetry was already put forward by the authors of Ref.~\cite{Bauer:2011ah} as a solution to the in $\epsilon_K$ problem in flavor anarchic scenarios. We analyze in  details the contributions to $\epsilon_K$ as well as $\epsilon'/\epsilon_K$ in anarchic scenarios based on the coset SU(3)$_L\times$SU(3)$_R/$SU(3)$_V$. The extended symmetry is broken in the UV regime and we assume this breaking is realized by appropriate boundary conditions on the UV brane. In order for SM fermion masses to arise the same symmetry must also be broken in the IR and we consider two distinct breaking mechanisms to achieve so. First of all, as in the main text of the paper, we assume existence of a bulk scalar $\phi$, singlet of the EW group and transforming as $({\bf 3,\bar 3})$ under SU(3)$_L\times$SU(3)$_R$. In that case, we demonstrate that no parametric improvement relative to the usual anarchic warped scenario is obtained  regarding CPV in $K-\bar K$ mixing. We then consider a case where the SM-like bulk Higgs scalar is charged under the extended color group and transforms as a $({\bf 3,\bar
3,2,\bar 2})_{0}$ of the full bulk gauge symmetry. By computing the NP contributions to $\epsilon_K$ and $\epsilon'/\epsilon_K$ arising in that case, we show that no significant improvement is obtained when both constraints are combined. 

\subsection{IR breaking of SU(3)$_L\times$SU(3)$_R$ from an EW singlet VEV}

NP contributions to $\epsilon_K$ arises at tree-level through FCNCs mediated by both the vector and axial KK-gluons. 
$\epsilon_K$ is dominated by the Wilson coefficient $C^4_K$, which is chirally enhanced and is estimated to be~\cite{APS,Bauer:2011ah}\footnote{$\gamma_f$ factors parameterizing the SM fermion to KK-gluons overlap are implicit.}
\beq\label{C4K}
C^4_K\sim g_V^2\chi_{Q^1}\chi_{D^1}\chi_{Q^2}\chi_{D^2}\left(\frac{1}{m_V^2}-\frac{1}{m_A^2}\right)\simeq g_V^2\chi_{Q^1}\chi_{D^1}\chi_{Q^2}\chi_{D^2}\frac{\delta m^2}{m_V^4}\,,
\eeq
with $\delta m^2\equiv m_A^2-m_V^2\ll m_V^2$ and $\chi_f$ is the value of the SM fermion $f$ profile on the IR brane. The $C^4_K$ contribution is suppressed relative the case with no extended color symmetry (obtained in the limit $m_A\to \infty$) provided the axial KK-gluon is degenerate with the vector KK-gluon, $\delta m^2\ll m_V^2$~\cite{Bauer:2011ah}. We argued in Appendix~\ref{splitting} that this splitting is constrained to be sizable by the large top mass, with $\delta m^2/m_V^2\gtrsim \mathcal{O}(0.1)$, thus limiting already the suppression in $C_4^K$.  Moreover we now show that this does not lead to an improvement on the associated bound on the KK scale, relative the anarchic RS model. All chirality flipping operators much involve an appropriate power of the scalar fiel $\phi$, due to the fermion embedding under the extended color group. In particular Yukawa interactions are suppressed by one power of $\phi$ VEV, such that $\epsilon_K$ contributions in Eq.~\eqref{C4K} is independent on  the spurion VEV $\langle\phi\rangle$ after trading the ``compositeness'' factors for the SM quark masses. We thus find 
\beq
C_K^4 \simeq \frac{2g_V^2}{m_V^2}\frac{m_dm_s}{Y^2v^2v_\phi^2}\frac{\delta m^2}{m_V^2}
\simeq R_A\times \left(C_K^4\right)_{\rm RS}\,,\quad {\rm where} \quad R_A\equiv \frac{g_V^2(t+t^{-1})\Lambda^2}{m_V^2}\left(\frac{Y_{\rm RS}}{Y}\right)^2
\eeq
and  $\left(C_K^4\right)_{\rm RS}$ is the $C_K^4$ expression found in anarchic RS model with SU(3)$_V$ color symmetry in the bulk~\cite{APS}. The size of $R_A$ is rather model dependent but it can be estimated from NDA. Assuming a bulk Higgs field we find the following NDA values for the 5D Yukawa couplings: $Y\sim 16\pi/\sqrt{N_c}$ and $Y_{\rm RS}\sim 4\pi/\sqrt{N_{\rm KK}}$. Noting that $t+t^{-1}\geq 2$, $g_V\gtrsim 3$~\cite{Agashe2site}, and using $\Lambda\sim N_{\rm KK}m_V$ we find $R_A\simeq 3N_{\rm KK}/4\sim \mathcal{O}(1)$ for a minimal value of $N_{\rm KK}=2$.  
We conclude that when the extended color symmetry is broken by an EW singlet scalar VEV, no parametric improvement relative to anarchic warped model with only one color SU(3) group is obtained in $\epsilon_K$.

\subsection{IR breaking of SU(3)$_L\times$SU(3)$_R$ from a colored Higgs VEV}

We now consider the case where the extended color symmetry is broken in the IR by a colored Higgs VEV.
We parameterize the  72 real components of $H\sim({\bf 2,\bar 2,3,\bar 3})_{0}$  as
\beq\label{Hcompo}
H-\langle H\rangle=(h+i\tilde h)\frac{\sigma^0\lambda^0}{\sqrt{12}}+(\chi_i+i\tilde\chi_i)\frac{\sigma^i\lambda^0}{\sqrt{12}}+
(\phi^a+i\tilde\phi^a)\frac{\sigma^0\lambda^a}{2\sqrt{2}}+(\psi_i^a+i\tilde\psi_i^a)\frac{\sigma^i\lambda^a}{\sqrt{12}}\,,
\eeq
where $i= 1,2,3$ and $a=1,\dots 8$, $\sigma^i$ and $\lambda^a$ are the Pauli and Gell-Mann
matrices, respectively, normalized such that $\tr(\sigma^i\sigma^j)=2\delta^{ij}$ and $\tr(\lambda^a\lambda^b)=2\delta^{ab}$, and $\sigma^0={\bf 1}_{2\times2}$ and $\lambda^0={\bf 1}_{3\times 3}$. The Higgs VEV is $\langle H \rangle = (v/\sqrt{12})\sigma^0\lambda^0$, with $v\simeq 246\,$GeV. Notice that $H$ no longer spans a pseudo-real irreducible representation due its color charges, so the Higgs field kinetic term is $\mathcal{L}_{\rm kin}^H=$tr$ |D^\mu H|^2$ and all components in Eq.~\eqref{Hcompo} are canonically normalized fields. There are eight color singlets and eight octets, half of each are electrically neutral.
The neutral and charged states will yield different flavor spurions, namely $Y_D^3$ and $Y_D^2 Y_U$, respectively. Under the flavor anarchy assumptions there is an unknown relative phase between these two flavor structures, so the neutral and charged contributions cannot be combined. Barring accidental cancellations, considering only one of the two contribution is a well representative and conservative choice. We choose to focus on the $Y_D^3$ contribution for simplicity. 

\subsubsection{Indirect CPV: $\epsilon_K$}

\begin{figure}
\begin{center}
\includegraphics[width=.6\textwidth]{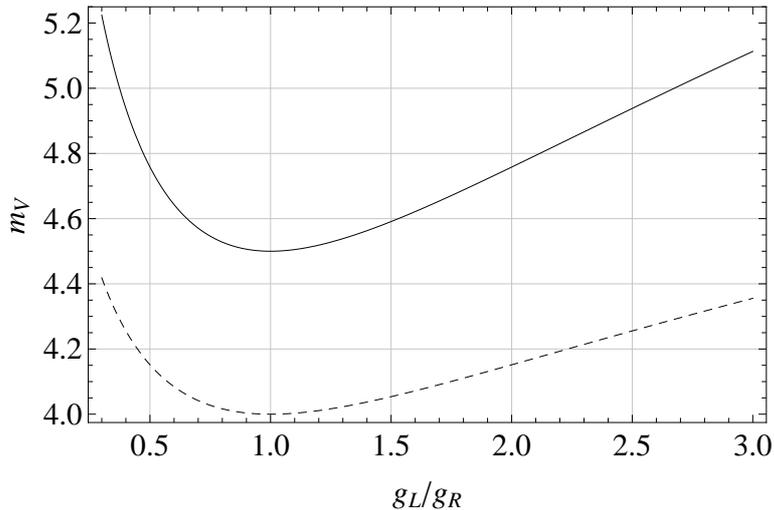}
\end{center}
\caption{Kaon physics' bounds on the KK scale $m_V$ as a function of $t\equiv g_L/g_R$ in flavor anarchic warped models based on the extended color symmetry SU(3)$_L\times$SU(3)$_R$. The continuous curve corresponds to the lowest $\epsilon_K$ bound for an IR-localized colored Higgs field, while the dashed curve shows the combined lowest bound from $\epsilon_K$ and $\epsilon'/\epsilon_K$ in the case of a bulk Higgs with $\beta=0$. The weakest bounds are obtained for $g_L=g_R$.} 
\label{anarchy-epsilon}
\end{figure}

We start with evaluating the tree-level contribution to $\epsilon_K$. The NP contribution to $C_K^4$ as expressed in Eq.~\eqref{C4K} still holds, but with a mass splitting of
\beq
\delta m^2 \simeq \frac{(g_L^2+g_R^2)}{6}v^2\,\mathcal{O}_A(\beta),
\eeq
where $\beta$ controls the localization of the Higgs VEV along the fifth dimension.
The resulting $C_K^4$ coefficient after the SM quark mass replacement is
\beq
\frac{C^4_K}{(C^4_K)_{\rm RS}}\simeq \frac{g_V^2(t+t^{-1})v^2\mathcal{O}_A(\beta)}{m_V^2}\left(\frac{Y_{\rm RS}}{Y}\right)^2\,,
\eeq
which is parametrically suppressed with respect to anarchic RS models.
Assuming an IR-brane localized Higgs the suppression factor is at most 
\beq
\left.\frac{C^4_K}{(C^{4}_K)_{\rm RS}}\right|_{IR}\gtrsim 0.5\times\left(\frac{2\,{\rm TeV}}{m_V}\right)^2\left(\frac{g_V}{3}\right)^2\,,
\eeq
where we assumed $Y\simeq Y_{\rm RS}$, as suggested by NDA. Following Ref.~\cite{GIP}, this translates into the following bound on the KK scale
\beq\label{eKIR}
m_V^{\epsilon_K}\big|_{\rm IR}\gtrsim 3.8\,{\rm TeV}\left(\frac{3}{Y}\right)^{1/2}\left(\frac{g_V}{3}\right)\,,
\eeq
where we assumed the Yukawa coupling saturates its perturbative value $Y\lesssim 2\pi/N_{KK}\simeq 3$ for $N_{\rm KK}=3$. 
Notice that the suppression factor is maximal for $t=1$; We show in Fig.~\ref{anarchy-epsilon} how the bound on the KK scale is worsened when moving away from this limit. 

As already observed in Ref.~\cite{Agashe2site}, flavor constraints are typically relaxed when the Higgs field propagates in the bulk thanks to a smaller overlap between the KK states and the Higgs VEV and a 5D Yukawa coupling. For a Higgs field maximally delocalized in the bulk ($\beta=0$), as arises in gauge-Higgs unification models, $C^4_K$ is further suppressed by a factor $\sim 2$ compared with the IR Higgs case above
\beq
\left.\frac{C^4_K}{C^{4\ RS}_K}\right|_{\rm bulk}\simeq 0.25 \left.\frac{C^4_K}{C^{4\ RS}_K}\right|_{\rm IR}\,.
\eeq
The above bound on the KK scale is relaxed to
\beq\label{eKbulk}
m_V^{\epsilon_K}\big|_{\rm bulk}\gtrsim \frac{5.2\,{\rm TeV}}{Y^{1/2}}\left(\frac{g_V}{3}\right)\simeq
1.9\ {\rm TeV}\left(\frac{7}{Y}\right)^{1/2}\left(\frac{g_V}{3}\right)\,,
\eeq
where $Y\sim 4\pi/\sqrt{N_{\rm KK}}\simeq 7$ is the maximal perturbative value for a bulk Higgs. The above bound from $\epsilon_K$ alone in the bulk Higgs case is weaker than that of EWPTs ($m_V\gtrsim 3\,$TeV typically); this is partially due to the extended color symmetry in the bulk. As observed in Ref.~\cite{GIP}, the large Yukawa coupling used to relax $\epsilon_K$ in the bulk Higgs case is potentially in tension with indirect CPV in Kaon decay. We thus discuss now contributions to $\epsilon'/\epsilon_K$. 

\subsubsection{Direct CPV: $\epsilon'/\epsilon_K$}

\begin{figure}[tb]
\begin{center}
\includegraphics[scale=0.7]{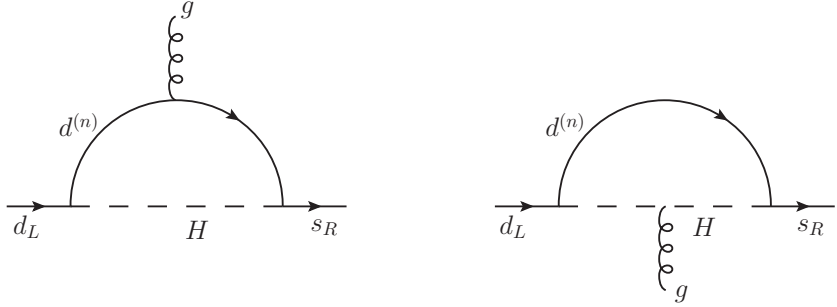}
\end{center}
\caption{One-loop diagrams contributing to the chromomagnetic operator $\mathcal{O}_G$ relevant to direct CP violation in Kaon decay. The required chirality flip is implicit and we only consider diagrams where down-type KK states $d^{(n)}$ yielding a $Y_D^3$ contribution run in the loops. The right-hand side diagram is absent for color singlet scalars.}
\label{fig-epsilon-prime}
\end{figure}

NP contributions to direct CPV in the $K^0\to\pi^+\pi^-$  decay channel arise from flavor violating chromomagnetic dipole operators like
\bea
\mathcal{O}_G&=&g_s\bar{s}_R\sigma_{\mu\nu}G^{\mu\nu} d_L\,,\\
\mathcal{O}_G'&=&g_s\bar{s}_L\sigma_{\mu\nu}G^{\mu\nu} d_R\,,
\eea
where $G^{\mu\nu}$ is the gluon field strength and $g_s$ the QCD gauge coupling. Since flavor violation is only mediated by the 5D anarchic Yukawa the dominant contribution to the Wilson coefficients $C_G$ and $C_G'$ arises at one-loop through diagrams where the Higgs states and KK-fermions run in the loop. We only consider contributions where KK-fermions of first KK level are running in the loops. As argued above we focus on one-loop contributions which only involve (electrically) neutral Higgs states.
When the Higgs field is charged under QCD we
find two one-loop diagrams which match onto the above operators, as shown in Fig.~\ref{fig-epsilon-prime}. We define $C_G=C_G^{(1)}+C_G^{(2)}$ and similarly $C_G'=C_G^{\prime\,(1)}+C_G^{\prime\,(2)}$, where the first (second) term denote the contribution from the left-hand (right-hand) side of Fig.~\ref{fig-epsilon-prime}.
The radiative contributions depend on the SM fermion embedding. We assume first that the LH and RH first two-generation SM fermions are respectively embedded in $({\bf 2,1})$ and $({\bf 1,2})$ representations of the EW custodial group SU(2)$_L\times$SU(2)$_R$; We will discuss other choices of representation below. Following Ref.~\cite{GIP}
we find the following contribution from the left-hand side diagram of Fig.~\ref{fig-epsilon-prime} after replacing the compositeness factors by appropriate SM masses and CKM mixings\footnote{We assumed the mass of the first KK fermions is the same of the KK scale, which is a good approximation in the anarchic case for KK partners of the first two generation quarks relevant to the loop amplitude considered.} 
\beq\label{CG1}
C_G^{(1)}\simeq -m_sV_{us}\frac{Y^2}{64\pi^2m_V^2}\mathcal{O}_\beta\left(\sum_{i={\rm singlets}}c_i - \sum_{i={\rm octets}} c_i\right)
\eeq
where first (second) term originates from color singlet (octet) scalar components and $c_i=(1,1,1/3,1/3)$ for $i=h,\chi_3,\tilde h,\tilde \chi_3$ and $i=\phi,\psi_3,\tilde \phi,\tilde \psi_3$. The $c_i$'s characterize the coupling strength of the scalar states to fermions relative to the SM-like Higgs ($h$) coupling. $\mathcal{O}_\beta$ parameterizes the Higgs VEV profile overlap with the KK state wavefunctions. As showed in Ref.~\cite{DelKam} it significantly depends on the Higgs field localization parameter $\beta$. For large values of $\beta\gtrsim \mathcal{O}(10)$ mimiking a Higgs field quasi-localized on the IR-brane, with a profile width $d\lesssim 1/m_V$ yet larger than the inverse 5D cut-off scale, $\mathcal{O}_\beta\sim \mathcal{O}(1)$. In contrast, for a maximally delocalized Higgs field ($\beta=0$) $\mathcal{O}_\beta\simeq 0.1$ leading to significant suppression of the dipole amplitude relative the IR Higgs case~\cite{DelKam}.  
$C_G^{\prime\,(1)}$ is obtained from the $C_G^{(1)}$ through the replacements $m_sV_{us}\to m_d/V_{us}$. We find that the singlet and octet states yield contribution of the same magnitude but with a relative sign, so that  $C_G^{(1)}=C_G^{\prime\,(1)}\simeq 0$. The extra minus sign orignates from the color contraction relevant to  the octet scalar contributions$\propto T^b T^a T^b=-T^a/2N_c$, while singlet contributions are simply $\propto T^a$. The factor of $2N_c$ difference between  octet and singlet contributions is compensated by the fact that octet couplings to fermions are $\sqrt{2N_c}$ larger than the singlet couplings, as it can be seen from Eq.~\eqref{Hcompo}.
Only color octet states contribute to the right-hand side diagram of Fig.~\ref{fig-epsilon-prime}. We find the following contribution to the Wilson coefficients:
\beq
C_G\simeq C_G^{(2)} \simeq m_sV_{us}\frac{9Y^2}{64\pi^2m_V^2}\mathcal{O}_\beta\sum_{i={\rm octets}}c_i=m_sV_{us}\frac{3Y^2}{8\pi^2m_V^2}\mathcal{O}_\beta\,,
\eeq
and $C_G^{\prime\,(2)}$ is obtained from $C_G^{(2)}$ through the replacement $m_sV_{us}\to m_d/V_{us}$. Only the $h$ and $\tilde\chi_3$ states in Eq.~\eqref{CG1} contribute to the anarchic RS result, yet with different couplings to fermions: $c_h^{\rm RS}=6$ and $c_{\tilde \chi_3}^{\rm RS}=2$. We thus find 
\beq\label{CGRS}
C_G^{\rm RS}\simeq -m_s V_{us}\frac{Y^2}{8\pi^2m_V^2}\mathcal{O}_\beta\,.
\eeq
We note that this result is a factor of 2/3 smaller than that of Ref.~\cite{GIP}. We thus find that the chromomagnetic dipole contribution to $\epsilon'/\epsilon_K$ arising from the colored Higgs is enhanced compared to the anarchic RS result of Eq.~\eqref{CGRS} by a factor of 3. Following Ref.~\cite{GIP}, this yields the following bounds for both the quasi-localized IR-Higgs and $\beta=0$ bulk Higgs cases
\beq\label{eprime}
m_{V}^{\epsilon'/\epsilon}\big|_{\rm IR}\gtrsim 5.4Y\,{\rm TeV}\,,\quad
m_{V}^{\epsilon'/\epsilon}\big|_{\rm bulk}\gtrsim1.7Y\,{\rm TeV}\,.
\eeq

We have moreover considered different sets of irreducible representations where all generations of SM quarks are embedded in $L/R$ symmetric representations of the EW custodial group~\cite{AgasheDarold}. We focused on two cases where the LH and RH down-type quarks are embedded in (a) $(\bf{2,\bar 2})$ and $({\bf3,1})+({\bf 1,3})$ and (b) $({\bf 2,\bar 2})$ and $({\bf 1,1})$ of SU(2)$_L\times$SU(2)$_R$. In both (a) and (b) cases, we found that the dipole contribution is enhanced by a factor of $(1+m_{\rm custo}/2m_V)$, where $m_{\rm custo}$ is the mass of the additional fermionic (custodian) states relative to the representations considered above.
Since $m_{\rm custo}\simeq m_V$ in anarchic RS, the presence of these extra KK fermion always enhances the chromomagnetic dipole contributions to $\epsilon'/\epsilon_K$ by a factor of $\simeq 1.5$.

\subsubsection{Combined direct and indirect CPV bounds}

As first pointed out in Ref.~\cite{GIP}, since contributions to $\epsilon_K$ and $\epsilon'/\epsilon_K$ scale differently with the 5D anarchic Yukawa, they should be jointly considered in order to bound the KK scale. Combining the indirect and direct CPV bounds derived in Eqs.~\eqref{eKIR}-\eqref{eKbulk} and~\eqref{eprime}) we find \bea
m_V^{\epsilon_K+\epsilon'/\epsilon}\big|_{\rm IR}&\gtrsim &6.2\ {\rm TeV} \left(\frac{g_V}{3}\right)^{2/3}\,,\\
m_V^{\epsilon_K+\epsilon'/\epsilon}\big|_{\rm bulk}&\gtrsim & 3.6\ {\rm TeV}\left(\frac{g_V}{3}\right)^{2/3}\,,
\eea
for the IR-Higgs and bulk Higgs ($\beta=0$) cases, respectively.
The above bounds are obtained upon optimizing for the 5D Yukawa in order to minimize both $\epsilon_K$ and $\epsilon'/\epsilon_K$ contributions; We find the optimal Yukawa to be
\beq
Y\big|_{\rm IR}\simeq 1.1\left(\frac{g_V}{3}\right)^{2/3}\,,\quad Y\big|_{\rm bulk}\simeq 2.1\left(\frac{g_V}{3}\right)^{2/3}\,.
\eeq
These results assume $t=1$. Moving away from this limit the bound
gets stronger, as displayed in Fig.~\ref{anarchy-epsilon}. The above bounds are to be compared with the corresponding combined bound for the anarchic RS model
\beq
(m_V^{\epsilon_K,\epsilon'/\epsilon}\big|_{\rm IR})_{\rm RS}\gtrsim 12\,{\rm TeV} \left(\frac{g_V}{3}\right)^{1/2}\,,\quad
(m_V^{\epsilon_K,\epsilon'/\epsilon}\big|_{\rm bulk})_{\rm RS}\gtrsim 5.0\,{\rm TeV} \left(\frac{g_V}{3}\right)^{1/2}\,.
\eeq
CPV flavor constraints in the Kaon system in models with an extended color bulk symmetry are relaxed relative the anarchic RS models. However the amelioration is mild ($ \sim 30\%$) for physically motivated bulk Higgs profile ($\beta=0$). We conclude that a CP problem remains in the SU(3)$_L\times$SU(3)$_R$ warped models with flavor anarchy.

\section{Colored Higgs mass fine-tunning}\label{finetuning}

Making the Higgs field a bi-fundamental of SU(3)$_L\times$SU(3)$_R$ introduces new scalar fields coupled to the SM-like Higgs boson. We show that these states significantly worsen the fine-tuning of the Higgs mass with respect to the SM case. We extract the quadratic divergences of the Higgs mass either by calculating the one-loop SM diagrams or using directly the Coleman-Weinberg (CW) potential~\cite{CW}. We begin by reviewing the SM calculation of the Higgs mass squared at one-loop.
Explicitly evaluating the SM one-loop diagrams yields the following quadratic divergences\footnote{We corrected for a missing  factor of 2 in the expression found in Ref.~\cite{Grojean:2007zz}.}
\beq\label{eqn:dmh2diag}
\delta m_h^2 = \frac{\Lambda^2}{16\pi^2}\left[6\lambda+\frac{1}{4}\left(9g^2+3g^{\prime2}\right)-6y_t^2\right]\,.
\eeq
The above result is gauge invariant\footnote{This is shown explicitly in calculating the loop amplitude in the $R_\xi$ gauge~\cite{Einhorn}. Since the momentum cut-off regularization procedure breaks gauge invariance, the result is  expected to be gauge dependent; the gauge independence of the above result may be a one-loop accident~\cite{Fukuda}.}. The use of the CW effective potential yields
\beq
\delta m_h^2 = \frac{\partial^2 V_{\rm eff}(h)}{\partial h^2}\,,
\eeq
with 
\beq
V_{\rm eff}(h) = \frac{1}{2}\int \frac{d^4k_E}{(2\pi)^4} {\rm Str} \log\left[k_E^2+M^2(h)\right]\,,
\eeq
where the super-trace Str runs over all fields whose mass depends on the background value of $h$. Evaluating the one-loop integral we find
\beq\label{eqn:dmh2CW}
\delta m_h^2 =\frac{\Lambda^2}{32\pi^2}\,{\rm Str}\frac{\partial^2 M^2(h)}{\partial h^2}\,.
\eeq
Using $m_W^2(h)=g^2h^2/4$, $m_Z^2(h)=(g^2+g^{\prime2})h^2/4$, $m_t^2(h)=y_t^2h^2/2$, $m_h^2(h)=\lambda (3h^2-v^2)$ and $m_\chi^2(h)=\lambda(h^2-v^2)$, Eq.~(\ref{eqn:dmh2CW}) reduces to 
Eq.~(\ref{eqn:dmh2diag}) when each contribution is weighted by the appropriate number of degrees of freedom. The latter being $n_W=2\times3=6$, $n_Z=3$, $n_t=3\times 4=12$, $n_h=1$ and $n_\chi=3$. This result is consistent with the diagram computation.
We now consider the extra quadratic divergences arising when the Higgs transforms as $H\sim (3,\bar{3},2,\bar 2)_0$. There are two distinct contributions arising from the axial KK gluon field and the additional scalar components (beyond the SM Higgs ones). We collectively denote the latter ($\tilde{h},\tilde{\chi}_i,\phi^a,\tilde{\phi}^a,\psi_i^a,\tilde{\psi}_i^a$) by $\phi$. The $h$-dependent masses are respectively
\bea
m_A^2(h)&\simeq& 0.96M_{\rm KK}^2+ (g_L^2+g_R^2)\frac{h^2}{6}\mathcal{O}_A(\beta)\,,\\
m_\phi^2(h)&=& M_\phi^2 + \lambda h^2
\eea
while $m_h^2(h)=\lambda(3h^2-2v^2)$ and $m_\chi^2(h)=\lambda(h^2-2v^2)$, as in the SM. $M_\phi^2$ denotes contributions from sources of SU(3)$_L\times$SU(3)$_R$ breaking other than the Higgs VEV. We made the implicit assumption that all scalar profiles are the same as that of the VEV, which should be a good enough approximation if all scalar masses are much smaller than $m_{V}$. Using the CW potential method one finds the following quadratic divergences in the custodial SU(3) model
\bea
\delta m_h^2 &=& \delta m_h^2\big|_{\rm SM} + \frac{\Lambda^2}{32\pi^2}\left[2\lambda\cdot (1+3)+2\lambda\cdot(2\times8)+2\lambda\cdot(2\times 3\times 8)\right]\,\nonumber\\
&&+\frac{\Lambda^2}{32\pi^2}(g_L^2+g_R^2)\frac{\mathcal{O}_A(\beta)}{3}\cdot(3\times8)\,\nonumber\\
&=&\delta m_h^2\big|_{\rm SM} + \frac{\Lambda^2}{16\pi^2}\left[68\lambda+4\left(g_L^2+g_R^2)\,\mathcal{O}_A(\beta)\right)\right]\,,\label{eqn:dmh2SU3}
\eea
where $\delta m_h^2\big|_{\rm SM}$ is given by Eq.~(\ref{eqn:dmh2diag}). In units of the scalar and gauge SM contributions, respectively, the two new contributions in Eq.~(\ref{eqn:dmh2SU3}) are 
\bea
\frac{\Lambda^2}{16\pi^2}\times68\lambda &=& \frac{34}{3}\,\delta m_h^2\big|_{\rm SM}^{h,\chi_i}\,,\\
\frac{\Lambda^2}{16\pi^2}\times4(g_L^2+g_R^2)\,\mathcal{O}_A(\beta) & = & \frac{16(g_L^2+g_R^2)\,\mathcal{O}_A(\beta)}{(9g^2+3g^{\prime2})}\, \delta m_h^2\big|_{\rm SM}^{W^\pm,Z}\,,\nonumber\\
&\gtrsim & \frac{64g_{V}^2\mathcal{O}_A(\beta)}{9g^2}\, \delta m_h^2\big|_{\rm SM}^{W^\pm,Z}\,,\nonumber\\
&\simeq& 150\left(\frac{g_{V}}{3}\right)^2\left(\frac{\mathcal{O}_A(\beta)}{1}\right)\,\delta m_h^2\big|_{\rm SM}^{W^\pm,Z}\,.
\eea
We conclude that taking the Higgs to transform as a $({\bf 3,\bar 3,2,\bar 2})_{0}$ worsens the fine-tuning by about one and two orders of magnitude in the scalar and gauge sector, respectively.

\end{document}